 \documentclass[12pt,a4paper]{article}

\usepackage{cite}
\usepackage{amsmath,amsfonts}
\usepackage{amssymb}
\usepackage[dvipdfmx]{graphicx}
\usepackage{color,comment, bm,authblk}

\usepackage{dcolumn}
\usepackage{bm}
\usepackage{aas_macros}
\usepackage{enumitem}
  \usepackage{tabularx}
   \newcolumntype{C}{>{\centering\arraybackslash}X}
   \newcolumntype{L}{>{\raggedright\arraybackslash}X}
   \newcolumntype{R}{>{\raggedleft\arraybackslash}X}
\setlistdepth{10}
\counterwithin{equation}{section}

\usepackage[normalem]{ulem}

\newcommand{\ii}{\mathrm{i}}
\newcommand{\dd}{\mathrm{d}}

\newcommand{\del}{\partial}
\newcommand{\ee}{{\rm e}}

\definecolor{DarkBlue}{rgb}{0,0,0.7} 

\definecolor{DarkRed}{rgb}{0.65,0,0}

\begin{document}


\author{Chul-Moon~Yoo\thanks{\sf yoo.chulmoon.k6@f.mail.nagoya-u.ac.jp}}

\affil{
Division of Particle and Astrophysical Science,
Graduate School of Science, Nagoya University, 
Nagoya 464-8602, Japan
}

\vskip-1.cm
\title{The basics of primordial black hole formation and \\abundance estimation}



\maketitle

\begin{abstract}
  \baselineskip5.5mm
  This paper is a biased review of primordial black hole (PBH) formation and abundance estimation. 
  We first review the three-zone model for PBH formation to 
  help an intuitive understanding of the PBH formation process. 
  Then, for more accurate analyses, we introduce necessary tools such as 
  cosmological long-wavelength solutions, the definition of the mass and compaction function in 
  a spherically symmetric spacetime and peak theory. 
  Combining all these tools, we calculate the PBH mass spectrum for the case of the monochromatic curvature power spectrum
  as a demonstration. 
  \end{abstract}
\clearpage
\tableofcontents
\clearpage

\section{Introduction}

Primordial black holes (PBHs) keep attracting the attention of researchers in cosmology, high energy and gravitational physics 
since they have been proposed~\cite{1967SvA....10..602Z,Hawking:1971ei,Carr:1974nx}.  
The specific black hole formation process is fascinating for people who are interested in 
general relativity and gravitational physics in the first place. 
The cosmological consequence can be sufficiently significant to make cosmologists pay attention to PBHs. 
Generation of the primordial fluctuations which may cause an abundant population of PBHs is closely related to the 
underlying inflationary scenario and the associated high-energy physics model.

In recent years, the possibility of PBHs has been seriously considered and observational constraints 
have been actively investigated. 
While we have not found evidence of the existence of PBHs, 
there are three interesting candidates: dark matter~\cite{Carr:2009jm,Carr:2020gox,Carr:2020xqk}, black hole binaries observed by gravitational wave interferometers~\cite{Bird:2016dcv,Sasaki:2016jop,Clesse:2017bsw,Sasaki:2018dmp} and 
the earth-mass objects of microlensing events~\cite{Niikura:2019kqi}. 
In addition, PBHs may explain a potential signal at a pulsar timing array~\cite{NANOGrav:2020bcs,Vaskonen:2020lbd,DeLuca:2020agl,Kohri:2020qqd,Sugiyama:2020roc,Domenech:2020ers,Inomata:2020xad} and the seeds of supermassive black holes and cosmic structures~\cite{Kawasaki:2012kn,Kohri:2014lza,Nakama:2016kfq,Carr:2018rid,Serpico:2020ehh,Unal:2020mts,Kohri:2022wzp}. 
It is very exciting to clarify whether those candidates are PBHs or not. 

There are several aspects of PBH research: early universe models which provide a substantial number of PBHs, 
dynamical process of PBH formation~\cite{Niemeyer:1997mt,Niemeyer:1999ak,Shibata:1999zs,2002CQGra..19.3687H,Musco:2004ak,Harada:2013epa,Nakama:2013ica,Nakama:2014fra,Harada:2015yda,Musco:2018rwt,Escriva:2019nsa,Escriva:2019phb,Escriva:2020tak,Escriva:2021aeh,Musco:2021sva,Escriva:2022bwe,Franciolini:2022tfm,Papanikolaou:2022cvo} effects of the evaporation of PBHs\cite{PhysRevD.19.1036,Barrow:1990he,Baumann:2007yr,Inomata:2020lmk,Papanikolaou:2020qtd,Domenech:2020ssp,Bhaumik:2020dor,Hooper:2020evu,Domenech:2021wkk}, associated gravitational wave background\cite{Ananda:2006af,Baumann:2007zm,Saito:2008jc,Saito:2009jt,Assadullahi:2009nf,Bugaev:2009zh,Bugaev:2010bb,Espinosa:2018eve,Kohri:2018awv,Domenech:2021ztg}, 
observational consequences and constraints~\cite{Carr:2020gox} and so on. 
The purpose of this review is to give a brief introduction to the PBH formation process and 
show a recently developed method for the abundance calculation with the PBH formation criterion and 
the peak statistics for the Gaussian curvature perturbation. 
  We would like to ask readers to refer to other recent reviews and lecture notes~(e.g., Refs.~\cite{Escriva:2022duf,Villanueva-Domingo:2021spv,Byrnes:2021jka,Sasaki:2018dmp}) on PBHs for the scientific significance and historical reviews.

In Sec.~\ref{sec:three_zone}, in order to provide an intuitive understanding of the PBH formation process, 
we introduce the three-zone model based on Ref.~\cite{Harada:2013epa}. 
The cosmological long-wavelength approximation, which is a necessary ingredient for PBH study, 
is reviewed in Sec.~\ref{sec:cosmo_long_sol} following Refs.~\cite{Shibata:1999zs,Harada:2015yda}. 
Assuming spherical symmetry, the PBH formation criterion is discussed in Sec.~\ref{sec:PBH_formation_cri} based on 
the compaction function introduced in Ref.~\cite{Shibata:1999zs} for the first time.  
Our method to estimate the PBH abundance in peak theory is reviewed in Sec.~\ref{sec:estPBH} 
following Refs.~\cite{Yoo:2018kvb,Yoo:2020dkz}. We calculate the PBH mass spectrum for the monochromatic curvature power spectrum 
as a demonstration. 
Brief reviews for a conserved mass in spherically symmetric spacetimes and peak theory~\cite{1986ApJ...304...15B} are attached as Appendices.

Throughout this paper, general relativity is assumed, and we use the geometrized units in which both 
the speed of light and Newton's gravitational constant are set to unity, $c=G=1$.

\section{PBH formation process: three-zone model}
\label{sec:three_zone}

First, in order to understand the overall picture of PBH formation, let us review 
the three-zone model discussed in Ref.~\cite{Harada:2013epa}. 
The three-zone model describes the collapsing system in a cosmological background with a patchwork 
of Friedmann-Lema\^itre-Robertson-Walker (FLRW) models. 

\subsection{FLRW models}
Let us write the line element of an FLRW model as follows:
\begin{equation}
  \dd s^2=-\dd t^2+a^2(t)\left[\frac{\dd r^2}{1-Kr^2}+r^2\dd \Omega^2\right], 
\end{equation}
where $a(t)$, $K$ and $\dd \Omega$ are the scale factor, spatial curvature and line element of the unit two-sphere. 
One of the Friedmann equations is given by 
\begin{equation}
  \left(\frac{\dot a}{a}\right)^2=\frac{8\pi}{3}\rho-\frac{K}{a^2}, 
  \label{eq:Hubble}
\end{equation}
where the dot ''$\dot~$" denotes the time derivative. 
Introducing the areal radius $R:=a(t)r$, Eq.~\eqref{eq:Hubble} can be 
rewritten as follows:
\begin{equation}
  \frac{\dot R^2}{2}-\frac{M}{R}=E(r):=-\frac{Kr^2}{2}, 
\end{equation}
where $M$ is the mass inside the comoving radius $r$ given by $4\pi\rho R^3/3$. 
This equation can be regarded as the energy conservation law for a 
fluid element on the comoving radius $r$. The first and second terms correspond to 
kinetic energy and gravitational potential energy, respectively. 
The time-independent total energy for the motion of the fluid element is given by the right-hand side. 
Since the second term is inversely 
proportional to the areal radius $R$, if $K>0$, 
initially expanding fluid will turn around at the maximum radius $R_{\rm max}=M/E(r)$. 
At the maximum expansion, the density $\rho_{\rm max}$ and the scale factor $a_{\rm max}$ satisfy
\begin{equation}
  \frac{8\pi}{3}\rho_{\rm max}=\frac{K}{a_{\rm max}^2}. 
\end{equation}
In contrast, the fluid continues to expand if $K\leq 0$. 

\subsection{Background flat FLRW}

We assume that the background universe is given by a spatially flat ($K=0$) FLRW universe.  
For the matter content, we suppose the perfect fluid with a linear equation of state given by 
$p=w\rho$ with a constant $w$, where $\rho$ and $p$ are the energy density and the pressure. 
The constant $w$ is given by $0$, $1/3$, and $-1$ for the matter-dominated (MD) universe, radiation-dominated (RD) universe, and a cosmological constant ($\Lambda$), respectively. 
From the conservation law of the fluid $\dd (\rho a^3)=-p\dd a^3$, 
we obtain 
\begin{equation}
  \rho\propto a^{-3(1+w)}. 
\end{equation}
Then, we find the behavior of the Hubble radius as 
\begin{equation}
  R_{\rm H}:=1/H \propto a^{3(1+w)/2}. 
\end{equation}

\begin{figure}[htbp]
  \begin{center}
  \includegraphics[scale=0.92]{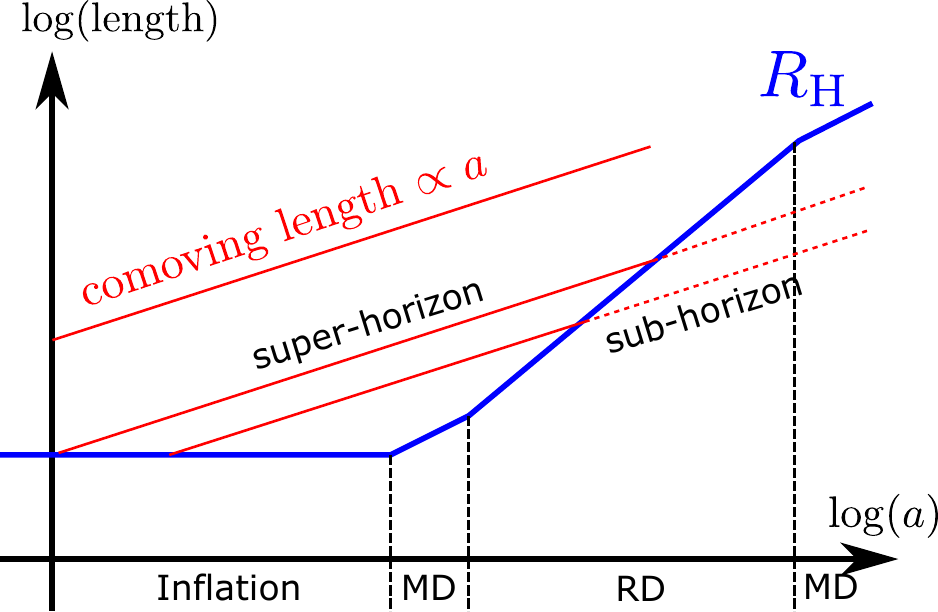}
  \end{center}
  \caption{\baselineskip5mm
  The rough sketch of the standard evolution of the background universe and 
the perturbations on it. 
  }\label{fig:length}
  \end{figure}

Let us review the rough sketch of the standard evolution of the background universe and 
the perturbations on it (see Fig.~\ref{fig:length}). 
In the standard scenario, an inflationary era ($R_{\rm H}\sim {\rm const.}$) is considered 
before the decelerated expansion phases of our universe. 
The inflationary era may be followed by an early MD phase ($R_{\rm H}\sim a^{3/2}$) supported by the inflaton oscillation, 
and the universe supposed to become RD universe ($R_{\rm H}\sim a^{2}$) after the completion of the reheating. 
During this early period, all relevant perturbations have super-horizon scales. 
As will be reviewed in subsequent sections, 
the amplitude of the curvature perturbation, 
which will be defined as the inhomogeneity of the spatial volume, 
is conserved on super-horizon scales. 
Thus the fluctuations generated in the inflationary era are ``frozen" after the horizon exit, 
and start to grow after the horizon entry.  

\subsection{Super-horizon inhomogeneity}

What we mainly discuss in this paper is the black hole formation after the horizon entry 
during the RD era. 
Therefore perturbations on super-horizon scales are considered for initial fluctuations 
of the subsequent evolution of an over-dense region. 
The evolution of super-horizon perturbations, whose comoving wavenumber $k$ satisfies $k\ll aH$, 
will be summarized in Sec.~\ref{sec:cosmo_long_sol}. 
Here let us take a result in advance. 
At the leading order of the long-wavelength perturbation ($\epsilon:=k/(aH)\ll1$), 
we obtain the following metric form: 
\begin{equation}
  \dd s^2=-\dd t^2+a^2\Psi(\bm x)\delta_{ij}\dd x^i \dd x^j, 
  \label{eq:leading}
\end{equation}
where $\delta_{ij}$ is the Kronecker delta and $\Psi(\bm x)$ is an arbitrary function of the spatial 
coordinates $x^i$. 
It would be fruitful to compare Eq.~\eqref{eq:leading} with the metric of the FLRW universe in 
the isotropic coordinates: 
\begin{equation}
  \dd s^2=-\dd t^2+a^2\frac{1}{\left(1+\frac{K}{4}|\bm x|^2\right)^2}\delta_{ij}\dd x^i\dd x^j. 
\end{equation}
At the leading order of the long-wavelength approximation, the conformal factor of the spatial metric can be 
an arbitrary function, that is, the long-wavelength inhomogeneity may be intuitively regarded as 
the position-dependent spatial curvature. 
Since a sufficiently overdense region is expected to have a moment of maximum expansion and start to contract, 
such a region may be considered to have locally positive curvature. 

\subsection{Three-zone model}

\begin{figure}[htbp]
\begin{center}
\includegraphics[width=10cm]{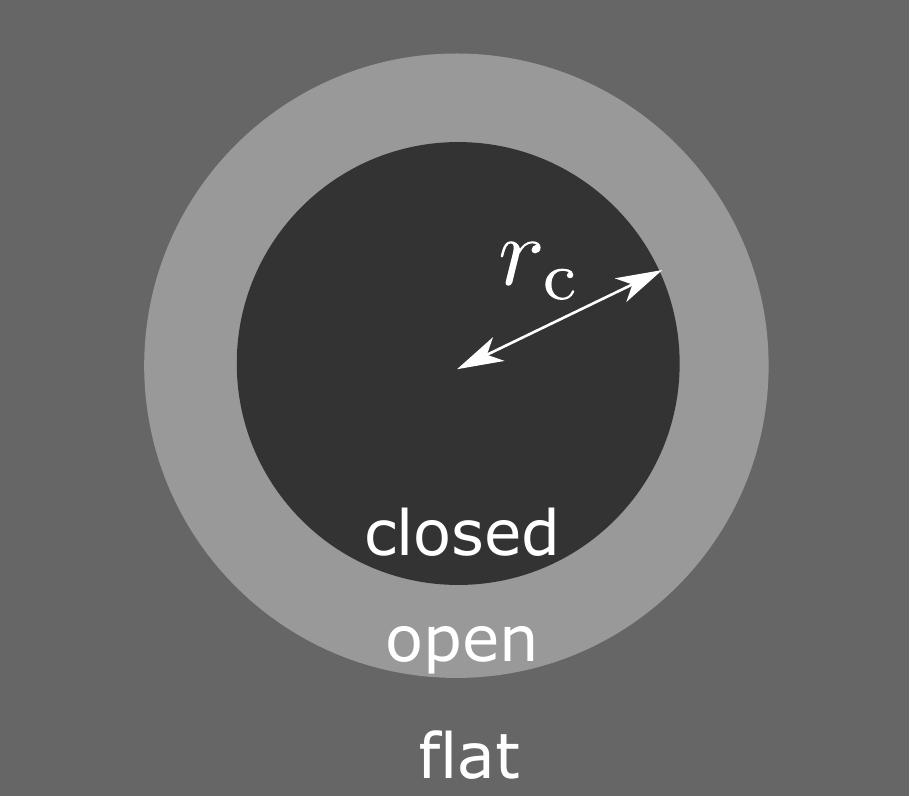}
\end{center}
\caption{Schematic figure of the three-zone model. }
\label{fig:3zone}
\end{figure}

Motivated by the previous subsection, let us consider the three-zone model composed of 
the spherical closed FLRW region $r<r_{\rm c}$ surrounded by an under-dense region embedded in the flat FLRW background (Fig.~\ref{fig:3zone}). 
This model can be an exact solution only for the dust case ($w=0$)\footnote{The under-dense region is necessary for the dust case to be an exact solution. }. 
In the case of $w\neq0$, the pressure gradient force diverges on the discontinuous spheres, and a singular shell 
must be introduced to realize the configuration as a solution of the Einstein equations. 
Nevertheless, the three-zone model is useful to intuitively understand the PBH formation process 
and roughly estimate the threshold for the amplitude of the density perturbation at the horizon entry. 

Let us take uniform Hubble time slices and compare the Friedmann equations \eqref{eq:Hubble} in the overdense region and the flat background. 
On a uniform Hubble time slice, since the value of the Hubble expansion rate is identical, 
we obtain 
\begin{equation}
  \frac{8}{3}\pi \rho_{\rm b}=  \frac{8}{3}\pi \rho-\frac{K}{a^2},
  \label{eq:uniH} 
\end{equation}
where $\rho$ and $\rho_{\rm b}$ denote the density in the overdense and background regions, respectively. 
The overdense region is described by the closed FLRW with the spatial curvature $K$. 
From Eq.~\eqref{eq:uniH}, the density perturbation can be calculated as 
\begin{equation}
  \delta:=\frac{\rho-\rho_{\rm b}}{\rho_{\rm b}}=\frac{K}{a^2H^2}. 
\end{equation}
The value of the density perturbation $\delta_{\rm H}$ at the horizon entry $R_{\rm c}:=ar_{\rm c}=1/H$ is given by 
\begin{equation}
  \delta_{\rm H}=Kr_{\rm c}^2.
  \label{eq:rdh}
\end{equation}
Since the value of the comoving radius $r$ cannot be larger than $\sqrt{K}$, we find the 
following geometrical upper bound~\cite{Kopp:2010sh}: 
\begin{equation}
  \delta_{\rm H}<1. 
\end{equation}

\subsection{Jeans criterion}

If the whole dynamics of the overdense region is described by the closed FLRW universe, 
nothing prevents the contraction of the closed universe and arbitrarily small amplitude of the positive density perturbation results in black hole formation. 
However, in reality, the pressure gradient would work as a preventing factor against the contraction. 
In order to estimate the effect of the pressure gradient, let us apply the Jeans criterion to the 
three-zone model. 
The Jeans criterion states that, if the free-fall time scale of the system 
is shorter than the sound-wave propagation time scale, the gravitational collapse cannot be prevented by 
the pressure gradient. 
For the overdense region, the sound-wave propagation time-scale $t_{\rm s}$ would be 
given by 
\begin{equation}
  t_{\rm s}=\frac{R_{\rm c}}{\sqrt{w}}=\frac{a r_{\rm c}}{\sqrt{w}}\propto a, 
  \label{eq:ts}
\end{equation}
where we have used the fact that the sound speed $c_{\rm s}$ is given by $\sqrt{w}$. 
The free-fall time-scale $t_{\rm ff}$ is given by 
\begin{equation}
  t_{\rm ff}=\frac{1}{H}=\left(\frac{8\pi}{3}\rho  \right)^{-1/2}=\frac{a_{\rm max}}{\sqrt{K}}\left(\frac{a}{a_{\rm max}}\right)^{3(1+w)/2}\propto a^{3(1+w)/2}. 
  \label{eq:tff}
\end{equation}
Let us define $a_{\rm th}$ as the value of the scale factor satisfying $t_{\rm s}|_{a=a_{\rm th}}=t_{\rm ff}|_{a=a_{\rm th}}$. 
Then if $a_{\rm max}>a_{\rm th}$, the sound-wave propagation time scale is shorter than the free-fall time scale at the time ${a=a_{\rm max}}$, and the gravitational contraction would be prevented by the pressure gradient effect. 
For the gravitational collapse and black hole formation to be realized, the following condition must be satisfied: 
\begin{equation}
  a_{\rm max}<a_{\rm th}=a_{\rm max}\left(\frac{w}{Kr_{\rm c}^2}\right)^{-1/(1+3w)}\Leftrightarrow Kr_{\rm c}^2=\delta_{\rm H}>w. 
  \label{eq:carrcon}
\end{equation}
This condition is often quoted as Carr's condition~\cite{Carr:1975qj}. 
Here it should be noted that Carr's condition has the ambiguity of a factor of $\mathcal O(1)$ 
coming from the order estimations of Eq.~\eqref{eq:ts} and Eq.~\eqref{eq:tff}. 

\subsection{Refinement of the threshold}

\begin{figure}[htbp]
\begin{center}
\includegraphics[width=10cm]{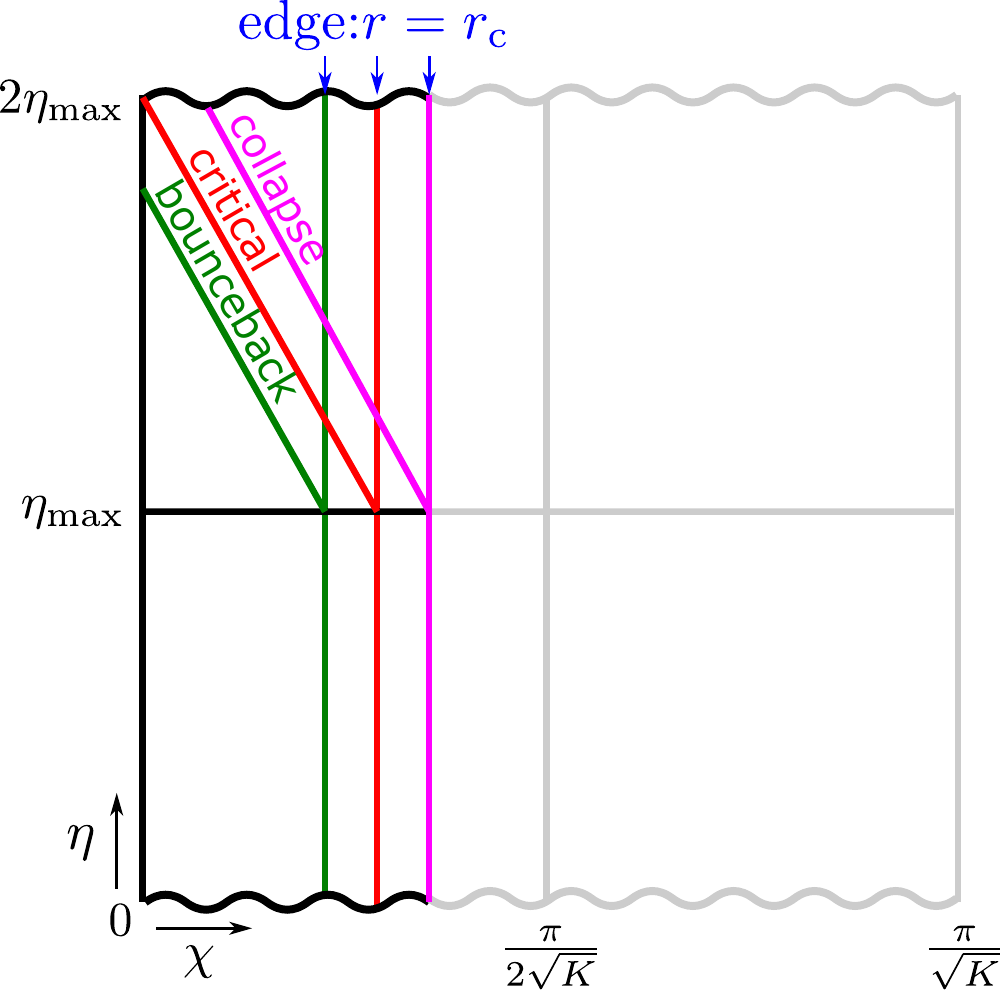}
\end{center}
\caption{Spacetime diagram for the closed universe with $w=1/3$ describing the overdense region and the trajectory of the sound wave propagation 
from the edge to the center emanated at the time of the maximum expansion. }
\label{fig:confd}
\end{figure}

Let us fix the ambiguous factor contained in Carr's condition \eqref{eq:carrcon}
based on a well-motivated way~\cite{Harada:2013epa} and make the condition more reliable. 
The basic idea is straightforward. 
We strictly define $t_{\rm s}$ and $t_{\rm ff}$ and explicitly calculate these time-scales 
based on the three-zone model. 
To do that, let us rewrite the metric of the overdense region in terms of the new radial coordinate $\chi$
defined by 
\begin{equation}
  \sin^2(\sqrt{K}\chi)=Kr^2  
\end{equation}
as follows: 
\begin{eqnarray}
\dd s^2&=&-\dd t^2+a^2\left[\dd \chi^2+\frac{1}{K}\sin^2(\sqrt{K}\chi)\dd\Omega^2\right] \\
&=&a^2\left[-\dd\eta^2+\dd \chi^2+\frac{1}{K}\sin^2(\sqrt{K}\chi)\dd\Omega^2\right],  
\end{eqnarray}
where we have also introduced the conformal time defined by $a\dd\eta=\dd t$. 
The Friedmann equation can be written as 
\begin{equation}
  \left[\frac{\dd}{\dd\eta}\ln a\right]^2=K\left[\left(\frac{a}{a_{\rm max}}\right)^{-1-3w}-1\right]. 
\end{equation}
Introducing $u:=(a/a_{\rm max})^{1+3w}$, we obtain
\begin{equation}
  \left(\frac{\dd u}{\dd \eta}\right)^2=K(1+3w)^2\left(u-u^2\right). 
\end{equation}
The solution is given by 
\begin{equation}
  u=\frac{1}{2}\left[1-\cos\left(\sqrt{K}(1+3w)\eta\right)\right], 
\end{equation}
where the value of $\eta$ is restricted to 
\begin{equation}
  0\leq\eta/2 \leq\eta_{\rm max}:=\frac{\pi}{\sqrt{K}(1+3w)}.
  \label{eq:etamax} 
\end{equation}
The times $\eta=0$ and $\eta=2\eta_{\rm max}$ correspond to the big-bang and big-crunch, 
and the maximum expansion is realized at $\eta=\eta_{\rm max}$. 
A spacetime diagram for the closed universe with $w=1/3$ is shown in Fig.~\ref{fig:confd}. 

Let us reinterpret the Jeans criterion. 
The propagation time scale of the sound wave can be defined as the time 
required for the sound wave to propagate from the edge of the overdense region to the center. 
The free-fall time scale can be simply replaced by the collapsing time of the closed universe describing the overdense region. 
Therefore if the sound wave emanated from the edge of the overdense region at the time of the maximum expansion 
reaches the center before the big crunch, we conclude that the gravitational collapse will be prevented by the pressure gradient effect. 

The sound speed $c_{\rm s}$ is given by $c_{\rm s}=\sqrt{w}$. 
Then the trajectory of the sound wave is parallel to the straight line $\chi =\pm \sqrt{w}\eta$. 
From Fig.~\ref{fig:confd}, we find that the Jeans criterion gives the threshold of the radius of the 
closed universe describing the overdense region for PBH formation.
This is quite reasonable because the edge radius $r_{\rm c}$ is related to the amplitude of the density perturbation 
at the horizon entry by Eq.~\eqref{eq:rdh}. 
The threshold value is calculated by the following equation:
\begin{equation}
  \sqrt{w}\eta_{\rm max}=\chi_{\rm c}, 
  \label{eq:forth}
\end{equation}
where $\chi_{\rm c}$ is the value of the coordinate $\chi$ at the edge of the overdense region, 
that is, defined by 
\begin{equation}
  \sin\left(\sqrt{K}\chi_{\rm c}\right)=\sqrt{K}r_{\rm c}. 
\label{eq:chic}
\end{equation}
Combining Eqs.~\eqref{eq:rdh}, \eqref{eq:etamax}, \eqref{eq:forth} and \eqref{eq:chic}, 
we can derive the following threshold value for the amplitude of the density perturbation at the horizon entry:
\begin{equation}
  \delta_{\rm th}=\sin^2\left(\frac{\pi \sqrt{w}}{1+3w}\right). 
\end{equation}
It should be noted that this value is for the density perturbation in the uniform Hubble gauge. 
In the comoving gauge, as will be shown later, the factor $3(1+w)/(5+3w)$ must be multiplied, namely, 
\begin{equation}
  \delta^{\rm co}_{\rm th}=\frac{3(1+w)}{(5+3w)}\sin^2\left(\frac{\pi \sqrt{w}}{1+3w}\right).   
  \label{eq:deltacoth}
\end{equation}

In this section, we have refined Carr's condition based on the three-zone model. 
However, our estimation does not go much beyond order estimation. 
For more rigorous quantitative estimation, we need to rely on numerical simulations, and 
the threshold value somewhat depends on the profile of the inhomogeneity.  
Nevertheless, it has been reported that the threshold \eqref{eq:deltacoth} approximately reproduces the lower 
bound of the threshold value~\cite{Musco:2018rwt}. 
Therefore, in terms of results, the three-zone model roughly provides the least possible effect 
of the pressure gradient against PBH formation. 

\section{Cosmological long-wavelength solutions}
\label{sec:cosmo_long_sol}
To investigate PBH formation more accurately, 
we need to investigate the inhomogeneities which are initially super-horizon scales. 
Such inhomogeneities are described by growing mode solutions in the long-wavelength approximation. 
The expansion of the long-wavelength approximation with the expansion parameter $\epsilon=k/(aH)$ 
is often called gradient expansion. 
In the context of PBH formation, there are two main formulations of gradient expansion. 
One is the gradient expansion of the equations in the Misner-Sharp formulation~\cite{Misner:1964je} in which 
the Einstein equations for a fluid system in spherical symmetry are written down with the comoving coordinates. 
The other is based on the cosmological 3+1 decomposition~\cite{Shibata:1999zs}. 
The relation between these two formulations is investigated in Ref.~\cite{Harada:2015yda}. 
Although the Misner-Sharp formulation has been adopted for PBH studies by many authors including Polnarev and Musco~\cite{Polnarev:2006aa}, 
in this paper, we review the formulation adopted in Ref.~\cite{Shibata:1999zs} because the latter allows more general gauge conditions. 
We follow and quote the discussions and calculations in Refs.~\cite{Shibata:1999zs,Harada:2015yda} in this section. 
Let us start with writing down the Einstein equations in the form of the cosmological 3+1 decomposition for a perfect fluid component. 
We just display the necessary equations for later discussions and ask readers to refer to Ref.~\cite{Harada:2015yda} for more details. 

\subsection{Cosmological 3+1 decomposition}
First, we write the line element of spacetime as follows:
\begin{equation}
  \dd s^2=-\alpha^2\dd t^2+a^2\psi^4\tilde \gamma_{ij}(\dd x^i+\beta^i\dd t)(\dd x^j+\beta^j \dd t),  
\end{equation}
where $a$ is the scale factor of the background universe, and the lapse function $\alpha$, shift vector $\beta^i$, 
spatial conformal factor $\psi$ and $\tilde \gamma_{ij}$ are functions of $t$ and $x^i$. 
We choose $\tilde \gamma_{ij}$ so that $\tilde \gamma :=\det(\tilde \gamma_{ij})=f:=\det(f_{ij})$, 
with  $f_{ij}$ being a time-independent metric of the flat three-space. 
The Latin indices run over 1 to 3 and we drop and raise the Latin indices of quantities without/with tilde by $\gamma_{ij}$/$\tilde\gamma_{ij}$ and 
$\gamma^{ij}$/$\tilde \gamma^{ij}$, respectively. 
The extrinsic curvature of a time slice can be decomposed as 
\begin{equation}
  K_{ij}=\psi^4a^2\left(\tilde A_{ij}+\frac{1}{3}\tilde \gamma_{ij}K\right), 
\end{equation}
where $\tilde A_{ij}$ satisfies 
\begin{equation}
  \tilde \gamma^{ij}\tilde A_{ij}=0. 
\end{equation}
Then the Hamiltonian and momentum constraint equations can be written as follows:
\begin{eqnarray}
&&  \tilde \triangle \psi=\frac{1}{8}\tilde {\mathcal R}^k_k\psi-2\pi \psi^5a^2E-\frac{1}{8}\psi^5 a^2\left(\tilde A_{ij}\tilde A^{ij}-\frac{2}{3}K^2\right), 
\label{eq:hamcon}\\
&&\tilde {\mathcal D}^j(\psi^6\tilde A_{ij})-\frac{2}{3}\psi^6\tilde {\mathcal D}_i K=8\pi J_i\psi^6, 
\label{eq:momcon}
\end{eqnarray}
where $E=n_\mu n_\nu T^{\mu\nu}$ and $J_i=-\gamma_{i\mu} n_\nu T^{\mu\nu}$ with 
$n_\mu$, $\gamma_{\mu\nu}$ and $T^{\mu\nu}$ being the 
unit-normal form of the time slice, spatial metric and stress-energy tensor, respectively. 
$\tilde \triangle$, $\tilde {\mathcal D}_i$ and $\tilde {\mathcal R}_{ij}$ are the Laplacian, covariant derivative and the Ricci tensor for $\tilde \gamma_{ij}$, respectively. 
The evolution equations for geometrical variables are given by 
\begin{eqnarray}
  (\partial_{t}-{\cal L}_{\beta})\tilde{\gamma}_{ij}&=&-2\alpha
   \tilde{A}_{ij}-\frac{2}{3}\tilde{\gamma}_{ij}{\cal D}_{k}\beta^{k}, 
 \label{eq:SS99_2.11}\\
  (\partial_{t}-{\cal L}_{\beta})\tilde{A}_{ij}&=&\frac{1}{a^{2}\psi^{4}}
 \left[\alpha\left({\cal R}_{ij}-\frac{\gamma_{ij}}{3}{\cal R}\right)
 -\left(D_{i}D_{j}\alpha-\frac{\gamma_{ij}}{3}D_{k}D^{k}\alpha\right)\right]
 \nonumber \\
 && +\alpha(K\tilde{A}_{ij}-2\tilde{A}_{ik}\tilde{A}_{j}^{k})
 -\frac{2}{3}({\cal D}_{k}\beta^{k})\tilde{A}_{ij}
 -\frac{8\pi\alpha}{a^{2}\psi^{4}}\left(S_{ij}-\frac{\gamma_{ij}}{3}S_{k}^{k}\right), 
 \label{eq:SS99_2.12}
 \\
 (\partial_{t}-{\cal L}_{\beta})\psi &=&
  -\frac{1}{2}H\psi+\frac{\psi}{6}(-\alpha K+{\cal D}_{k}\beta^{k}), \label{eq:SS99_2.13} \\
 (\partial_{t}-{\cal L}_{\beta})K &=&
  \alpha\left(\tilde{A}_{ij}\tilde{A}^{ij}+\frac{1}{3}K^{2}\right)-D_{k}D^{k}\alpha
  +4\pi\alpha (E+S_{k}^{k}), 
 \label{eq:SS99_2.14}
 \end{eqnarray}
  where $D_i$ and ${\mathcal D}_i$ are the covariant derivatives for $\gamma_{ij}$ and the 3-dim flat metric $f_{ij}$, respectively, 
  and $\mathcal R_{ij}$ is the Ricci tensor for $\gamma_{ij}$. 
  $S_{ij}$ is defined by $\gamma_{i\mu}\gamma_{j\nu}T^{\mu\nu}$.

For a perfect fluid, 
the stress-energy tensor can be written as 
\begin{equation}
  T_{\mu\nu}=(\rho +p)u_{\mu}u_{\nu}+p g_{\mu\nu},
 \label{eq:perfect_fluid}
 \end{equation} 
 where $u^{\mu}$ is the four-velocity of the fluid element which is 
 normalized as $u^{\mu}u_{\mu}=-1$. 
 Introducing 
 \begin{equation}
  v^{l}:=\frac{u^{l}}{u^{t}}
 \label{eq:v^l}
 \end{equation}
and 
\begin{equation}
  \Gamma:=\alpha u^{t}=\left[1-\alpha^{-2}(\beta_{k}+v_{k})(\beta^{k}+v^{k})\right]^{-1/2}, 
  \label{eq:w}
  \end{equation}
we can write the hydrodynamical equations in the form 
\begin{eqnarray}
  &&\left[\psi^{6}a^{3}\left\{(\rho+p)\Gamma^{2}-p\right\}\right]_{,t}
  +\frac{1}{\sqrt{f}}\left[\sqrt{f}
  \psi^{6}a^{3}\left\{(\rho+p)\Gamma^{2}-p\right\}v^{l}\right]_{,l}
  \nonumber \\
  &=&-\frac{1}{\sqrt{f}}\left[\sqrt{f}\psi^{6}a^{3}p(v^{l}+\beta^{l})\right]_{,l}
  +\alpha\psi^{6}a^{3}pK
  -\alpha^{-1}\alpha_{,l}\psi^{6}a^{3}\Gamma^{2}(\rho+p)(v^{l}+\beta^{l})
  \nonumber \\
  &&
   +\alpha^{-1}\psi^{10}a^{5}\Gamma^{2}(\rho+p)(v^{l}+\beta^{l})(v^{m}+\beta^{m})\left(\tilde{A}_{lm}+\frac{\tilde{\gamma}_{lm}}{3}K\right), 
  \label{eq:Energy_eq_decomposition}\\
  &&(\Gamma\psi^{6}a^{3}(\rho+p)u_{j})_{,t}+\frac{1}{\sqrt{f}}(\sqrt{f}
  \Gamma\psi^{6}a^{3}(\rho+p)v^{k}u_{j})_{,k} \nonumber \\
  &=&-\alpha\psi^{6}a^{3}p_{,j}+\Gamma\psi^{6}a^{3}(\rho+p)
  \left(-\Gamma\alpha_{,j}+u_{k}\beta^{k}_{,j}-\frac{u_{k}u_{l}}{2u^{t}}\gamma^{kl}_{,j}\right),
  \label{eq:SS99_2.9}
  \end{eqnarray}
  and
  \begin{eqnarray}
  (\Gamma\psi^{6}a^{3}n)_{,t}+\frac{1}{\sqrt{f}}
  (\sqrt{f}\Gamma\psi^{6}a^{3}nv^{k})_{,k}=0.
  \label{eq:SS99_2.8}
  \end{eqnarray}

\subsection{Gradient expansion}
In practice, the gradient expansion is performed by replacing the spatial derivative $\del_i$ by $\epsilon \del_i$ 
with a fictitious parameter $\epsilon$. Expanding the equations of motion in a power series of $\epsilon$, 
finally, we set $\epsilon=1$. 
Letting $k$ be the comoving wavenumber for the scale of inhomogeneity of interest, we may consider $|\del_i|\sim k$. 
Since the only natural scale of the system other than $k$ is $H$, the fictitious parameter can be regarded as 
$\epsilon\sim k/(aH)$. 
We also have to make key assumptions for the leading order $\epsilon\rightarrow 0$ to perform the gradient expansion. 
At the leading order, we assume that the universe is given by a spatially flat FLRW universe. 
More specifically, we require $\alpha-1=\mathcal O(\epsilon)$, $\beta^i=\mathcal O(\epsilon)$ and $\del_t \tilde\gamma_{ij}=\mathcal O(\epsilon)$. 
In addition, we assume that $\psi$ is identically unity somewhere in the universe, so that $a(t)$ is 
the scale factor for that part of the universe. 
Under these assumptions, it has been shown that $\del_t \tilde \gamma_{ij}$ temporally decays and 
one can set $\del_t \tilde\gamma_{ij}=\mathcal O(\epsilon^2)$ for growing mode solutions~\cite{Lyth:2004gb}. 

In this review, we mainly focus on two specific time slicing conditions: the constant mean curvature (CMC) slicing ($K=-3H$) and comoving slicing ($n^\mu=u^\mu$). 
Readers may refer to Ref.~\cite{Harada:2015yda} for other gauge conditions. 
The comoving slicing implies $u_i=u^t(v_i+\beta_i)=0$, then $\Gamma=1$, $E=\rho$ and 
$J_i=(\rho+p)u_i=0$. 
Since $\tilde A_{ij}=\mathcal O(\epsilon^2)$ from Eq.~\eqref{eq:SS99_2.11}, from the  momentum \eqref{eq:momcon} and Hamiltonian \eqref{eq:hamcon} 
constraints,  
we obtain $\mathcal {\tilde D}_i K=\mathcal O(\epsilon^3)$ and $H^2=\frac{8\pi}{3}\rho+\mathcal O(\epsilon^2)$. 
Therefore CMC, uniform density and comoving slicings coincide to $\mathcal O(\epsilon)$. 
In addition, from Eq.~\eqref{eq:Energy_eq_decomposition}, we obtain 
\begin{equation}
  2\del_t \ln \psi=H-\del_t \ln a+\mathcal O(\epsilon^2), 
\end{equation}
where we have assumed that the pressure is homogeneous to $\mathcal O(\epsilon)$. 
Since the right-hand side is a function of $t$, $\del_t \ln \psi$ is also homogeneous to $\mathcal O(\epsilon)$. 
Since we have assumed that $\psi$ is identically unity at some point, we conclude 
\begin{equation}
  \del_t \psi=\mathcal O(\epsilon^2). 
\end{equation}

Coming back to the general expressions without specifying the slicing condition,  
from $\del_t\psi=\mathcal O(\epsilon^2)$ and the equations of motion displayed in the previous section, 
one can find 
\begin{eqnarray}
  &&\psi=\mathcal O(\epsilon^0), ~v^i=\mathcal O(\epsilon), ~\rho=\rho_{\rm b}(1+\mathcal O(\epsilon^2)),~
  \tilde A_{ij}=\mathcal O(\epsilon^2),\cr
  &&\tilde \gamma_{ij}=f_{ij}+\mathcal O(\epsilon^2),~\alpha=1+\mathcal O(\epsilon^2),~
  K=3H(1+\mathcal O(\epsilon^2)), 
\end{eqnarray}
where $\rho_{\rm b}$ is the background density. 
Then we introduce the following perturbation variables:
\begin{eqnarray}
\psi(t,\bm x)&=&\Psi(\bm x)(1+\xi(t,\bm x))+\mathcal O(\epsilon^3),\\
\tilde \gamma_{ij}(t,\bm x)&=&f_{ij}+h_{ij}(t,\bm x)+\mathcal O(\epsilon^3),
\label{eq:hij}
\\
K(t,\bm x)&=&3H(1+\kappa(t,\bm x))+\mathcal O(\epsilon^3),\\
\alpha(t,\bm x)&=&1+\chi(t,\bm x)+\mathcal O(\epsilon^3),\\
\beta^i(t,\bm x)&=&\mathcal O(\epsilon),\\
\rho(t,\bm x)&=&\rho_{\rm b}(1+\delta(t,\bm x))+\mathcal O(\epsilon^3),\\
v^i(t,\bm x)&=&\mathcal O(\epsilon),
\end{eqnarray}
where $\Psi(\bm x)=\mathcal O(\epsilon^0)$ and $\xi$, $h_{ij}$, $\kappa$, $\chi$ and $\delta$ are the perturbation variables of $\mathcal O(\epsilon^2)$. 

Hereafter, we focus on the case of the linear equation of state $p=w\rho$ and consider the normal threading $\beta^i=0$ for simplicity. 
Readers may refer to Ref.~\cite{Harada:2015yda} for other general cases. 
The equations of motion can be reduced to the following:
\begin{eqnarray}
&&  \del_t \delta+3H(1+w)(\chi+\kappa)=O(\epsilon^{4}),
\label{eq:deltadot}\\
&&\partial_{t}[a^{3}(1+w)\rho_{\rm b}u_{j}]=-a^{3}\rho_{\rm b}
\left[w\partial_{j}\delta+(1+w)\partial_{j}\chi\right]+O(\epsilon^{5}).
\label{eq:SS99_3.3_variant}, \\
&& 6\del_t{\xi }-3H(\chi+\kappa)
=O(\epsilon^{4}), 
\label{eq:SS99_3.4} \\
&&
\triangle \Psi=-2\pi \Psi^{5}a^{2}\rho_{\rm b}(\delta-2\kappa)+O(\epsilon^{4}),
\label{eq:SS99_3.5}
\\
&& H^{-1}\del_t{\kappa} = 
\frac{1}{2}(3w-1)\kappa-\frac{3}{2}(1+w)\chi-\frac{1}{2}\delta(1+3w)+O(\epsilon^{4}),
\label{eq:SS99_3.12} \\
&& \partial_{t}h_{ij}=-2\tilde{A}_{ij}
+O(\epsilon^{4}), 
\label{eq:SS99_3.7}\\
&& \partial_{t}\tilde{A}_{ij}+3H\tilde{A}_{ij}=\frac{1}{a^{2}\Psi^{4}}
\Bigl[-\frac{2}{\Psi}\left({\cal D}_{i}{\cal D}_{j}\Psi-\frac{1}{3}f_{ij}{\triangle}\Psi\right)\cr
&&\hspace{3cm}+\frac{6}{\Psi^{2}}\left({\cal D}_{i}\Psi{\cal D}_{j}\Psi-\frac{1}{3}f_{ij}{\cal D}^{k}\Psi{\cal D}_{k}\Psi\right)\Bigr]+O(\epsilon^{4}),
\label{eq:SS99_3.8}\\
&& {\cal D}_{i}(\Psi^{6}\tilde{A}^{i}_{j})+2H\Psi^{6}{\cal D}_{j}\kappa=8\pi\Psi^{6}(1+w)\rho_{\rm b}u_{j}
+O(\epsilon^{5}), 
\label{eq:momentum_constraint_CMC_approx}
\end{eqnarray}

The solutions for Eqs.~\eqref{eq:SS99_3.7} and \eqref{eq:SS99_3.8} are given by 
\begin{eqnarray}
  h_{ij}&=&-\frac{4}{(3w+5)(3w-1)}p_{ij}\left(\frac{1}{aH}\right)^{2}+O(\epsilon^{4}), 
 \label{eq:h_ij_solution}\\
 \tilde{A}_{ij}&=&
 \frac{2}{3w+5}p_{ij}H
 \left(\frac{1}{aH}\right)^{2}+O(\epsilon^{4}), 
 \label{eq:A_tilde_ij_solution}
 \end{eqnarray}
respectively, where 
\begin{equation}
  p_{ij}(\bm x):=\frac{1}{\Psi^{4}}
 \left[
 -\frac{2}{\Psi}\left({\cal D}_{i}{\cal D}_{j}\Psi-\frac{1}{3}f_{ij}\triangle\Psi\right)+\frac{6}{\Psi^{2}}\left({\cal D}_{i}\Psi{\cal D}_{j}\Psi-\frac{1}{3}f_{ij}{\cal D}^{k}\Psi{\cal D}_{k}\Psi\right)
 \right]
 \end{equation}
 with $\triangle:=f^{ij}\mathcal D_i\mathcal D_j$.

 To find solutions for other quantities, we need to fix the slicing condition. 
 Let us consider CMC slicing ($\kappa=0$). 
 From Eq.~\eqref{eq:SS99_3.5}, we obtain
 \begin{equation}
  \delta=q\left(\frac{1}{aH}\right)^2+\mathcal O(\epsilon^4), 
  \label{eq:deltaCMC}
 \end{equation}
 where 
 \begin{equation}
  q(\bm x)=-\frac{4}{3}\frac{\triangle \Psi}{\Psi}. 
 \end{equation}
Then, from Eqs.~\eqref{eq:SS99_3.12}, \eqref{eq:SS99_3.4} and \eqref{eq:SS99_3.3_variant}, $\chi$, $\xi$ and $u_j$ are given as 
\begin{eqnarray}
  \chi&=&-\frac{3w+1}{3(1+w)}q\left(\frac{1}{aH}\right)^2+\mathcal O(\epsilon^4),\\ 
  \xi&=&-\frac{1}{6(1+w)}q\left(\frac{1}{aH}\right)^2+\mathcal O(\epsilon^4),\\
  u_j&=&v_j+\mathcal O(\epsilon^5)=\frac{2}{3(1+w)(3w+5)}\del_j q a\left(\frac{1}{aH}\right)^3+\mathcal O(\epsilon^5). 
\end{eqnarray}
One can check that the solutions solve the other equations not used to derive the solutions. 

For the comoving slicing ($u_j=0$), 
from Eq.~\eqref{eq:SS99_3.3_variant}, we obtain 
\begin{equation}
  \chi=\frac{w}{1+w}\delta+\mathcal O(\epsilon^4). 
\end{equation}
Substituting this relation into Eq.~\eqref{eq:deltadot} and \eqref{eq:SS99_3.12}, and combining them, 
we can derive the following solutions for $\delta$ and $\kappa$:
\begin{eqnarray}
  \delta&=&\frac{3(1+w)}{3w+5}q\left(\frac{1}{aH}\right)^2+\mathcal O(\epsilon^4),\\
  \kappa&=&-\frac{1}{3w+5}q\left(\frac{1}{aH}\right)^2+\mathcal O(\epsilon^4). 
\end{eqnarray}
Then $\chi$ can be given as 
\begin{equation}
  \chi=-\frac{3}{3w+5}q\left(\frac{1}{aH}\right)^2+\mathcal O(\epsilon^4). 
\end{equation}
From Eq.~\eqref{eq:SS99_3.4}, we obtain 
\begin{equation}
  \xi=-\frac{1}{2(3w+5)}q\left(\frac{1}{aH}\right)^2+\mathcal O(\epsilon^4). 
\end{equation}

Since the function $\Psi(\bm x)$, and therefore $q(\bm x)$ are shared in the both gauge conditions, 
we can easily find the relation of gauge-dependent quantities. 
For instance, we find 
\begin{equation}
  \delta_{\rm co}=\frac{3(1+w)}{3w+5}\delta_{\rm CMC}+\mathcal O(\epsilon^4), 
\end{equation}
where $\delta_{\rm CMC}$ and $\delta_{\rm co}$ are density perturbations for CMC and comoving gauges, respectively. 

\section{PBH formation criterion}
\label{sec:PBH_formation_cri}

In this section, we consider spherically symmetric systems. 
In earlier days, the amplitude of the density perturbation averaged within a certain radius is used to give  
an analytic PBH formation criterion for convenience. 
More specifically, letting $\bar \delta$ be the amplitude of the volume average of the density perturbation at the horizon entry, 
if $\bar \delta$ exceeds a certain threshold value, we think that PBH formation is realized. 
The threshold value has been estimated analytically and numerically by many authors~\cite{Carr:1975qj,1978SvA....22..129N,1980SvA....24..147N,Shibata:1999zs,Niemeyer:1999ak,Musco:2004ak,Polnarev:2006aa,Musco:2008hv,Musco:2012au,Harada:2013epa,Nakama:2013ica,Nakama:2014fra}. 

A recent trend is to use the ``compaction function'' proposed in Ref.~\cite{Shibata:1999zs}. 
There are several efforts to refine the simple treatment of the compaction function for a more accurate and convenient criterion of PBH formation~\cite{Escriva:2019nsa,Escriva:2019phb,Escriva:2020tak,Musco:2018rwt}.  
In this review, however, to avoid complicated discussions, 
we do not go into details about the recent refinements, but just review the simple use of the compaction function. 
That is, if the maximum value of the compaction function exceeds a certain threshold value, we think that PBH formation is realized. 
In this procedure, the threshold value depends on the profile of the initial perturbation. 
Nevertheless, we expect that, for moderate profiles, the deviation due to the profile dependence would be about 10\%. 
Regarding the compaction function, recently, we found a misunderstanding that has not been realized for a long time~\cite{Harada:2023ffo}. 
This misunderstanding will be also resolved. 

Let us start with defining Shibata-Sasaki's original compaction function in terms of the cosmological long-wavelength perturbations as follows: 
\begin{equation}
  \mathcal C_{\rm SS}(r):=\frac{\delta M_{\rm SS}}{R}~~{\rm with}~~\delta M_{\rm SS}:=4\pi a^3 \rho_{\rm b}\int^r_0\dd x \delta_{\rm CMC} (\Psi^2 x)^2\frac{\dd}{\dd x}(\Psi^2 x), 
  \label{eq:CSS}
\end{equation}
where $R$ is the areal radius of the sphere of the coordinate radius $r$ given by $R=ar\Psi^2$. 
By using the solution \eqref{eq:deltaCMC}, one can derive the following expression~\cite{Harada:2015yda}:
\begin{equation}
  \mathcal C_{\rm SS}(r)=\frac{1}{2}\left[1-\left(1+2r\frac{\dd}{\dd r}\ln\Psi\right)^2\right]. 
\end{equation}
From this expression, we find that $\mathcal C_{\rm SS}$ is time independent and 
calculated by the gauge independent quantity $\Psi$. 
Furthermore, $\Psi$ is related to the curvature perturbation, 
which is often used in the context of cosmological perturbation in the early universe, 
as $\zeta=-\frac{1}{2}\ln \Psi$
\footnote{We choose the sign convention so that the spatial metric will be given by $a^2\ee^{-2\zeta}\tilde \gamma_{ij}\dd x^i\dd x^j$. }. 
Let us introduce the radius $r_{\rm m}$ at which the compaction function takes the maximum value $\mathcal C_{\rm SS}^{\rm max}=\mathcal C_{\rm SS}(r_{\rm m})$. 
Then, we adopt the following PBH formation criterion: 
\begin{equation}
  \mathcal C^{\rm max}_{\rm SS}\gtrsim \mathcal C_{\rm SS}^{\rm th} 
  \label{eq:PBHcriterion}
\end{equation}
with the uncertainty coming from the profile dependence. 
As is shown in Figs.~2 and 3 in Ref.~\cite{Harada:2015yda} for two specific one-parameter families of initial profiles, 
the value of $\mathcal C_{\rm th}$ can be roughly estimated as $0.4$ with about 10\% uncertainty. 
For the specific profile given by $\ln\Psi \propto \sin(r)/r$, which corresponds to the typical profile for the monochromatic power spectrum of $\ln\Psi$, 
we obtain $\mathcal C_{\rm th}\simeq 0.44$ from numerical simulations. 
Therefore we use the value $\mathcal C_{\rm th}=0.44$ as a reference value in this paper. 

In order to understand the compaction function more deeply, let us review the relation between the compaction function and 
the conserved mass. 
In spherical symmetry, we have a well-behaved conserved mass called Misner-Sharp mass~\cite{Misner:1964je}. 
As is shown in the Appendix, the Misner-Sharp mass is equivalent to the Kodama mass defined by 
the charge associated with a conserved current. 
In addition, the value of the Misner-Sharp mass on a marginally trapped surface is 
given by half of the areal radius of the marginally trapped surface. 
Therefore the Misner-Sharp mass is closely related to horizon formation, and 
relevant to be used for horizon formation criterion. 


From the expression of the Kodama current and mass given in Eqs.~\eqref{eq:Kcur} and \eqref{eq:Kodama_mass}, 
we can derive the following expression:
\begin{eqnarray}
  M(t,r)&=&4\pi\int^r_0\dd x x^2 a^3 \alpha \psi^6 T^t_{~\mu} K^\mu\cr
  &=&4\pi a^3\int^r_0\dd x x^2 \psi^4\left\{-[(\rho +p)u^tu_t+p]\del_x(\psi^2 x)+(\rho+p)u^tu_r \frac{x}{a}\del_t(\psi^2a)\right\}. 
  \label{eq:Kodama_expand}
\end{eqnarray}
In the comoving slice ($u_r=0$), we obtain 
\begin{eqnarray}
  M_{\rm co}(t,r)&=&4\pi a^3\int^r_0\dd x x^2 \psi^4\rho\left\{\del_x(\psi^2 x)+\mathcal O(\epsilon^4)\right\}. 
\end{eqnarray}
The mass excess compared to the background FLRW with the identical areal radius is calculated as 
\begin{equation}
  \delta M_{\rm co}:=M(t,r)-M_{\rm FLRW}(t,\psi^2 r)=4\pi a^3 \rho_{\rm b}\int^r_0\dd x\left[(\Psi^2 x)^2\del_x(\Psi^2 x) \delta_{\rm co}+\mathcal O(\epsilon^4)\right]. 
\end{equation}
Then we can define the compaction function in the comoving gauge as 
\begin{equation}
  \mathcal C_{\rm co}:=\frac{\delta M_{\rm co}}{R}=\frac{3(1+w)}{3w+5}\frac{1}{2}\left[1-(1+2r\del_r \ln\Psi)^2\right]=\frac{3(1+w)}{3w+5}\mathcal C_{\rm SS}, 
\end{equation}
where $R=a\Psi^2 r$. 
Therefore $\mathcal C_{\rm SS}$ is the compaction function in the comoving gauge multiplied by $\frac{3(1+w)}{3w+5}$. 
However, it should be noted that, for the CMC slice, the contribution from the second term in the integrand of Eq.~\eqref{eq:Kodama_expand} 
cannot be neglected\footnote{In Refs.~\cite{Shibata:1999zs,Harada:2015yda}, the second term in the integrand of Eq.~\eqref{eq:Kodama_expand} 
has been erroneously dropped, and the authors got the wrong relation $\mathcal C_{\rm CMC}=\mathcal C_{\rm SS}$. 
The original definition of $\mathcal C_{\rm SS}$ given in Eq.~\eqref{eq:CSS} would have been provided based on this wrong calculation neglecting the difference between the CMC and comoving slice conditions. }  and we find $\mathcal C_{\rm CMC}\neq \mathcal C_{\rm SS}$. 
For a more concrete description, readers may refer to Ref.~\cite{Harada:2023ffo}. 

To see the relation between the compaction function and the volume average of the density perturbation,  
let us define the volume average of the density perturbation in the comoving slice as follows:
\begin{equation}
  \bar \delta:=\frac{4\pi \int^r_0 \dd x (\Psi^2 x)^2 \del_x (\Psi^2 x)\delta_{\rm co}}{4\pi (\Psi^2 r)^3/3}. 
\end{equation}
Then we can easily find 
\begin{equation}
  \bar \delta=\frac{2}{H^2R^2}\frac{\delta M_{\rm co}}{R}. 
\end{equation}
Therefore we obtain 
\begin{equation}
  \left.\bar \delta\right|_{HR=1}=2\mathcal C_{\rm co}. 
\end{equation}
That is, the compaction function can be also regarded as the amplitude of the volume average of the density perturbation at the horizon entry. 
This relation also supports the relevance of the compaction function to be used for the criterion of PBH formation. 

\section{Estimation of PBH abundance in peak theory}
\label{sec:estPBH}

We start with writing the spatial metric in the following form:
\begin{equation}
  \dd s_3^2=a^2\ee^{-2\zeta}\tilde \gamma_{ij}\dd x^i \dd x^j. 
\end{equation}
The curvature perturbation $\zeta$ is related to $\Psi$ as 
\begin{equation}
  \zeta=-\frac{1}{2}\ln \Psi. 
\end{equation}
Here we assume that $\zeta$ is the Gaussian random variable with the 
power spectrum $\mathcal P(k)$ defined by
\begin{equation}
  \langle\tilde \zeta^*(\bm k)\tilde \zeta(\bm k')\rangle=\frac{2\pi^2}{k^3}\mathcal P(k)(2\pi)^3\delta(\bm k-\bm k')
\end{equation}
with $\tilde \zeta(\bm k)=\int \dd^3 x \zeta(\bm x)\ee^{\ii \bm k \bm x}$. 
Readers may refer to Refs.~\cite{Yoo:2019pma,Kitajima:2021fpq,Escriva:2022pnz} for non-Gaussian cases. 
From the random Gaussian assumption and the gradient moments
$\sigma_n$ given by 
\begin{equation}
  \sigma_n^2=\int \dd \ln k k^{2n}\mathcal P(k), 
\end{equation}
as is shown in Appendices \ref{sec:number_density}, \ref{sec:typical_profile} and \ref{sec:peaksoflapzeta}, 
we can obtain the peak number density $n_{\rm pk}^{(k_\bullet)}(\tilde \mu_2,k_\bullet)$ and the typical peak profile $\bar \zeta(r)$ 
characterized by the two parameters $\tilde \mu_2$ and $k_\bullet$ defined by 
\begin{eqnarray}
  \tilde \mu_2&=&\triangle \zeta(0) \frac{\sigma_1^2}{\sigma_2^2}, \\
  k_\bullet^2&=&-\frac{\triangle\triangle\zeta(0)}{\tilde \mu_2}\frac{\sigma_1^2}{\sigma_2^2},   
\end{eqnarray}
where $\triangle$ is the flat Laplacian. 
Here we just write down the final expressions \eqref{eq:nkbullet} and \eqref{eq:zetabar}: 
\begin{eqnarray}
   n^{(k_\bullet)}_{\rm pk}(\tilde \mu_2, k_\bullet) d \tilde \mu_2 d k_\bullet
  =\frac{2 }{3^{3/2}(2\pi)^{3/2}}\tilde \mu_2 k_\bullet\frac{\sigma_2^3\sigma_4^2}{\sigma_1^4\sigma_3^3}f\left(\tilde \mu_2 k_\bullet^2\frac{\sigma_2^2}{\sigma_1^2\sigma_4}\right)
  P_1\left(\tilde \mu_2\frac{\sigma_2}{\sigma_1^2},\tilde \mu_2 k_\bullet^2\frac{\sigma_2^2}{\sigma_1^2\sigma_4};\gamma_3\right) 
  \dd\tilde \mu_2 \dd k_\bullet, &&
  \label{eq:nkbullet-t}\\
    \frac{\overline{\zeta}(r)}{\tilde \mu_2}=-\frac{1}{1-\gamma_3^2}\left(\psi_1+\frac{1}{3}R_3^2\triangle\psi_1\right)+k_\bullet^2\frac{1}{\gamma_3(1-\gamma_3^2)}\frac{\sigma_2}{\sigma_4}\left(\gamma_3^2\psi_1+\frac{1}{3}R_3^2\triangle\psi_1 \right), 
    ~&&
    \label{eq:zetabar-t}
\end{eqnarray}
where $P_1$, $f$, $R_n$ and $\psi_n$ are defined in Eqs.~\eqref{eq:p1}, \eqref{eq:funcf}, \eqref{eq:Rn} and \eqref{eq:psin}, respectively, and we omitted the last term in \eqref{eq:bar_delta_zeta}, which can be absorbed into the renormalization of the scale factor.

\begin{figure}[htbp]
\begin{center}
\includegraphics[width=12cm]{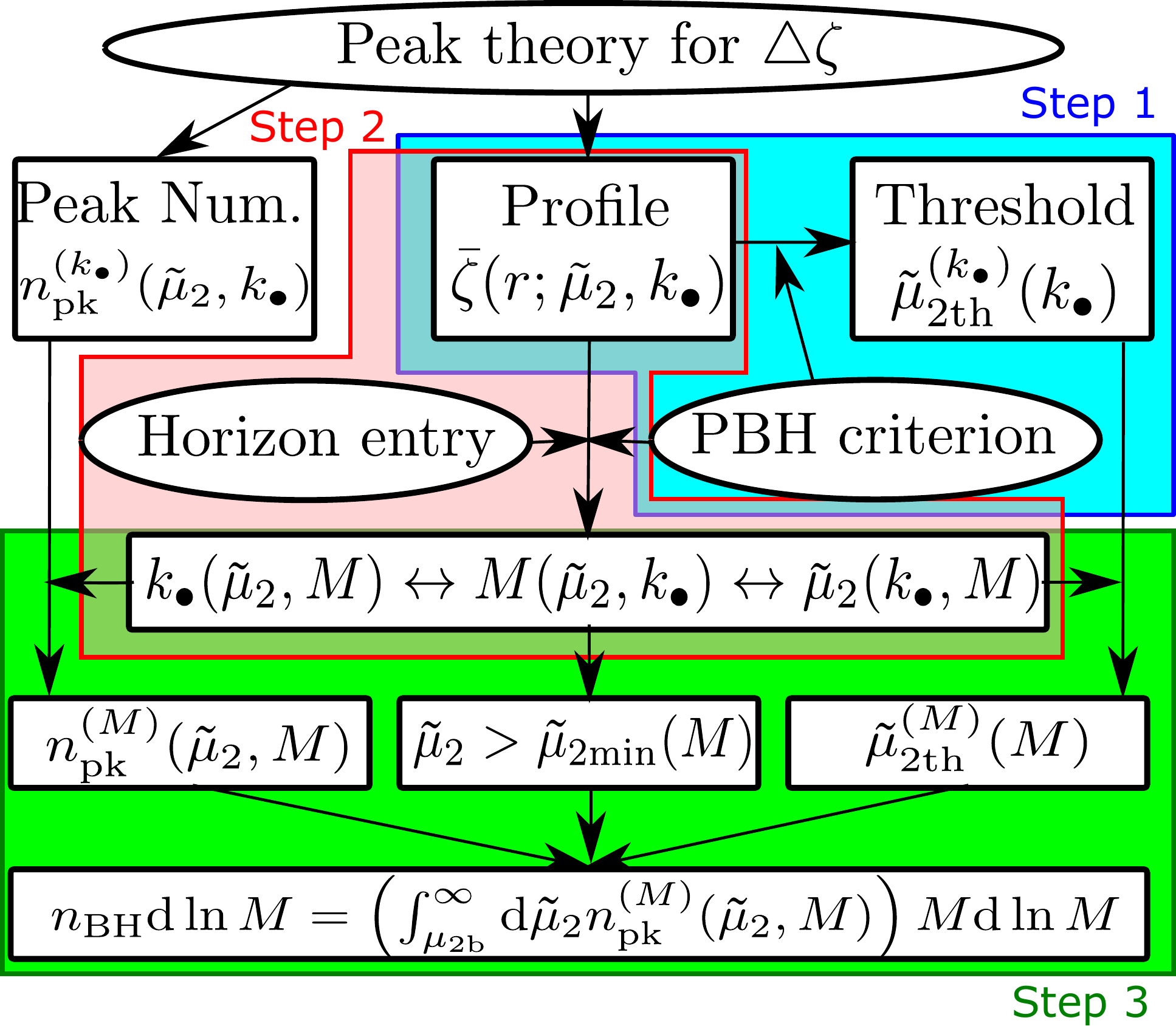}
\end{center}
\caption{A flow chart to reach the PBH mass spectrum. }
\label{fig:flow}
\end{figure}

A flow chart to reach the PBH mass spectrum by using the peak number density \eqref{eq:nkbullet-t}, typical profile \eqref{eq:zetabar-t} and the PBH formation criterion \eqref{eq:PBHcriterion} is shown in Fig.~\ref{fig:flow}. Let us explain each step of the flow chart. 

\subsection{Step 1: Rewriting the criterion }
First, let us consider the compaction function for the typical profile. 
Hereafter we will work in the comoving gauge and express the compaction function in the comoving gauge without the subscript, that is, 
$\mathcal C=\mathcal C_{\rm co}$. 
Focusing on the radiation-dominated universe ($w=1/3$), the PBH formation criterion can be rewritten as 
\begin{equation}
  \mathcal C^{\rm max}=\mathcal C(r_{\rm m})=\frac{2}{3}\mathcal C^{\rm max}_{\rm SS}\gtrsim \mathcal C^{\rm th}\sim 0.293
  \label{eq:Cth}
\end{equation}
In terms of the typical profile $\bar \zeta$, the compaction function can be written as 
\begin{equation}
  \mathcal C=\frac{1}{3}\left[1-\left(1-r\del_r\bar \zeta\right)^2\right]. 
\end{equation}
Assuming type I fluctuations defined by $\del_r R>0\Rightarrow \del_r (\ee^{-\zeta}r)>0$, 
we obtain 
\begin{equation}
  \left(\del_r \bar \zeta+r\del_r^2\zeta\right)|_{r=r_{\rm m}}=0. 
  \label{eq:eqrm}
\end{equation}
Defining the function $g(r;k_\bullet)$ by 
\begin{equation}
  g(r;k_\bullet):=\frac{\bar \zeta(r)}{\tilde \mu_2}, 
\end{equation}
the condition \eqref{eq:eqrm} for $r_{\rm m}$ can be rewritten as 
\begin{equation}
  (\del_r g+r \del_r^2 g)|_{r=r_{\rm m}}=0. 
\end{equation}
This equation implicitly gives the value of $r_{\rm m}$ as a function of $k_\bullet$: $r_{\rm m}=r_{\rm m}(k_\bullet)$. 
The value $\mathcal C^{\rm max}$ is then given as a function of $\tilde \mu_2$ and $k_\bullet$ through $r_{\rm m}(k_\bullet)$ as 
\begin{equation}
  \tilde \mu_2=\frac{1-\sqrt{1-3\mathcal C_{\rm max}}}{r_{\rm m}\del_r g(r_{\rm m})}. 
\end{equation}
Through this equation, we can obtain the threshold value of $\tilde \mu_2$ as a function of $k_\bullet$ as 
\begin{equation}
  \tilde \mu_{\rm 2th}^{(k_\bullet)}(k_\bullet):=\frac{1-\sqrt{\mathcal C_{\rm th}}}{r_{\rm m}(k_\bullet)\del_r g_{\rm m}(k_\bullet)}, 
  \label{eq:muth_k}
\end{equation}
where 
$\del_r g_{\rm m}=\del_r g(r_{\rm m}(k_\bullet);k_\bullet)$.

\subsection{Step 2: Derivation of the PBH mass expression }
In order to obtain the number density for a fixed PBH mass, we need to know the relation between the parameters $\tilde \mu_2$, $k_\bullet$ and the PBH mass $M$. 
The typical mass of a PBH can be roughly estimated by the horizon mass of $1/(2H)$ at the horizon entry. 
The horizon entry condition is given by 
\begin{equation}
  aH=\frac{a}{R}=\frac{\ee^{\tilde \mu_2 g_{\rm m}}}{r_{\rm m}}. 
  \label{eq:horizon_entry}
\end{equation}
In the radiation-dominated universe, adopting the matter-radiation equality as a reference time, we find 
\begin{equation}
  a=a_{\rm eq}\left(\frac{H_{\rm eq}}{H}\right)^{1/2}. 
  \label{eq:aHrad}
\end{equation}
From Eqs.~\eqref{eq:horizon_entry} and \eqref{eq:aHrad}, at the horizon entry, we obtain 
\begin{equation}
  H^{-1}=a_{\rm eq}^2H_{\rm eq}r_{\rm m}^2\ee^{-2\tilde \mu_2 g_{\rm m}}=H_{\rm eq}^{-1}k_{\rm eq}^2r_{\rm m}^2\ee^{-2\tilde \mu_2g_{\rm m}}, 
\end{equation}
where $k_{\rm eq}=a_{\rm eq}H_{\rm eq}$. 
Then the PBH mass can be estimated as 
\begin{equation}
  M=\frac{1}{2H}=\frac{1}{2}H_{\rm eq}^{-1}k_{\rm eq}^2r_{\rm m}^2\ee^{-2\tilde \mu_2g_{\rm m}}=M_{\rm eq}k_{\rm eq}^2r_{\rm m}^2\ee^{-2\tilde \mu_2g_{\rm m}},
  \label{eq:PBHMass}
\end{equation}
where we have introduced $M_{\rm eq}=1/(2H_{\rm eq})$. 
We have not taken into account the mass scaling associated with the critical behavior~\cite{Choptuik:1992jv,Koike:1995jm} in the expression \eqref{eq:PBHMass}. 
It should be also noted that the mass estimation given in this paper cannot be applied to Type II PBH formation~\cite{Kopp:2010sh}. 

\subsection{Step 3: Derivation of the mass function}
Eq.~\eqref{eq:PBHMass} gives us the relation between $M$, $k_\bullet$ and $\tilde \mu_2$. 
Therefore we can change the independent variables in the peak number density \eqref{eq:nkbullet-t} from $\tilde \mu_2$ and $k_\bullet$ to $\tilde \mu_2$ and $M$ 
as follows:
\begin{eqnarray}
    n^{(M)}_{\rm pk}(\tilde \mu_2, M) d \tilde \mu_2 d \ln M
   &=&\frac{2 }{3^{3/2}(2\pi)^{3/2}}\tilde \mu_2 k_\bullet\frac{\sigma_2^3\sigma_4^2}{\sigma_1^4\sigma_3^3}f\left(\tilde \mu_2 k_\bullet^2\frac{\sigma_2^2}{\sigma_1^2\sigma_4}\right)\cr
   &&
   P_1\left(\tilde \mu_2\frac{\sigma_2}{\sigma_1^2},\tilde \mu_2 k_\bullet^2\frac{\sigma_2^2}{\sigma_1^2\sigma_4};\gamma_3\right) 
   \left|\frac{\dd \ln M}{\dd k_\bullet}\right|^{-1}
   \dd \tilde \mu_2 \dd \ln M, 
   \label{eq:nM}   
\end{eqnarray}
where $k_\bullet$ should be regarded as a function of $\tilde \mu_2$ and $M$ through Eq.~\eqref{eq:PBHMass}. 
To obtain the PBH number density for a given value of the mass $M$, 
we need to integrate Eq.~\eqref{eq:nM} for $\tilde \mu_2$. 

As is explicitly shown in Ref.~\cite{Yoo:2020dkz} for some specific cases, 
the threshold value of $\tilde \mu_2$ given by Eq.~\eqref{eq:muth_k} can be converted to the 
threshold value $\tilde \mu_{\rm 2th}^{(M)}(M)$ for a given value of $M$ by using Eqs.~\eqref{eq:muth_k} and \eqref{eq:PBHMass}. 
In addition, the value of $\tilde \mu_2$ may be bounded below for a fixed value of $M$. 
Then, letting the minimum possible value of $\tilde \mu_2$ be $\tilde \mu_{\rm 2min}(M)$, the lower limit of the integral for $\tilde \mu_2$ is given by 
$\tilde \mu_{\rm 2b}:=\max \left\{\tilde \mu_{\rm 2min}(M),\tilde \mu_{\rm 2th}^{(M)}(M)\right\}$. 
That is, the PBH number density $n_{\rm BH}(M)$ and the mass spectrum $f_0(M)$ at equality time are given by 
\begin{eqnarray}
  n_{\rm BH}\dd \ln M&=&\left[\int^\infty_{\tilde \mu_{\rm 2b}}\dd\tilde \mu_2 n^{(M)}_{\rm pk}(\tilde \mu_2,M)\right]\dd \ln M,\\
  f_0\dd \ln M&=&\frac{M n_{\rm BH}}{\rho a^3}\frac{a_{\rm eq}}{a}\dd \ln M=\frac{4\pi}{3}\frac{M}{M_{\rm eq}}k_{\rm eq}^{-3}n_{\rm BH}\dd \ln M. 
\end{eqnarray}

\subsection{Critical behavior}
It is well known that, at least for idealized spherically-symmetric situations, the mass of the black hole 
scales as $\propto (b-b_{\rm th})^\gamma$ with $b$ and $b_{\rm th}$ being one parameter characterizing the initial data and the threshold value of it, respectively. 
PBH formation is no exception, and its realization and the effects on the PBH abundance have been investigated~\cite{Niemeyer:1999ak,Musco:2004ak,Musco:2008hv,Musco:2012au,Niemeyer:1997mt,Yokoyama:1998xd,Green:1999xm,Kuhnel:2015vtw,Germani:2018jgr,Biagetti:2021eep,Kitajima:2021fpq}. 
If we take the critical behavior into account, the PBH mass would be given by 
\begin{equation}
  M=K(k_\bullet)(\tilde \mu_2-\tilde \mu_{\rm 2th}^{(k_\bullet)}(k_\bullet))^\gamma M_{\rm eq} k_{\rm eq}^2r_{\rm m}^2\ee^{-2\tilde \mu_2 g_{\rm m}}, 
  \label{eq:PBHMass_crt}
\end{equation}
where $\gamma$ is the universal exponent given by $\gamma\simeq0.36$ for radiation fluid, 
and $K(k_\bullet)$ is a profile-dependent numerical factor, which will be set to be 1 hereafter for simplicity. 
In practice, we can consider the case without critical behavior setting the value of $\gamma$ to be 0 in Eq.~\eqref{eq:PBHMass_crt}. 

\subsection{The monochromatic curvature power spectrum}
In this paper, we show only the case of the monochromatic curvature power spectrum. 
Readers may refer to Refs.~\cite{Yoo:2018kvb,Yoo:2020dkz} for extended spectra. 
Let us write the monochromatic power spectrum as 
\begin{equation}
  \mathcal P(k)=\sigma_0^2k_0\delta(k-k0)=\sigma_0\delta(\ln k-\ln k_0). 
\end{equation}
Then the gradient moments are calculated as 
\begin{equation}
  \sigma_n^2=\sigma_0^2k_0^{2n}\Rightarrow\gamma_n=1. 
\end{equation}
The typical profile is given as 
\begin{equation}
  g(r;k_\bullet)=\frac{\bar \zeta}{\tilde \mu_2}=-\psi_1=-\frac{\sin(k_0 r)}{k_0r}. 
\end{equation}
From the condition \eqref{eq:eqrm}, we can find 
\begin{equation}
  r_{\rm m}\simeq 2.74/k_0,~g_{\rm m}=g(r_{\rm m})\simeq-0.141. 
\end{equation}
From the condition \eqref{eq:Cth}, we can find 
\begin{equation}
  \tilde \mu_{\rm 2th}\simeq0.615. 
\end{equation}
Then the PBH mass can be given by a function of $\tilde \mu_2$ as 
\begin{equation}
  M=(\tilde \mu_2-\tilde \mu_{\rm 2th})^\gamma M_{\rm eq}k_{\rm eq}^2r_{\rm m}^2\ee^{-2\tilde \mu_2 g_{\rm m}}, 
  \label{eq:Mmu}
\end{equation}
where $\gamma=0.36$ and $0$ with and without taking into account the critical behavior, respectively.
The minimum value of the PBH mass $M_{\rm min}$ is given by $0$ and $M_{\rm eq}k_{\rm eq}^2r_{\rm m}^2 \ee^{-2\tilde \mu_{\rm 2th}g_{\rm m}}$ for $\gamma=0.36$ and $0$, respectively. 

\begin{figure}[htbp]
\begin{center}
\includegraphics[width=12cm]{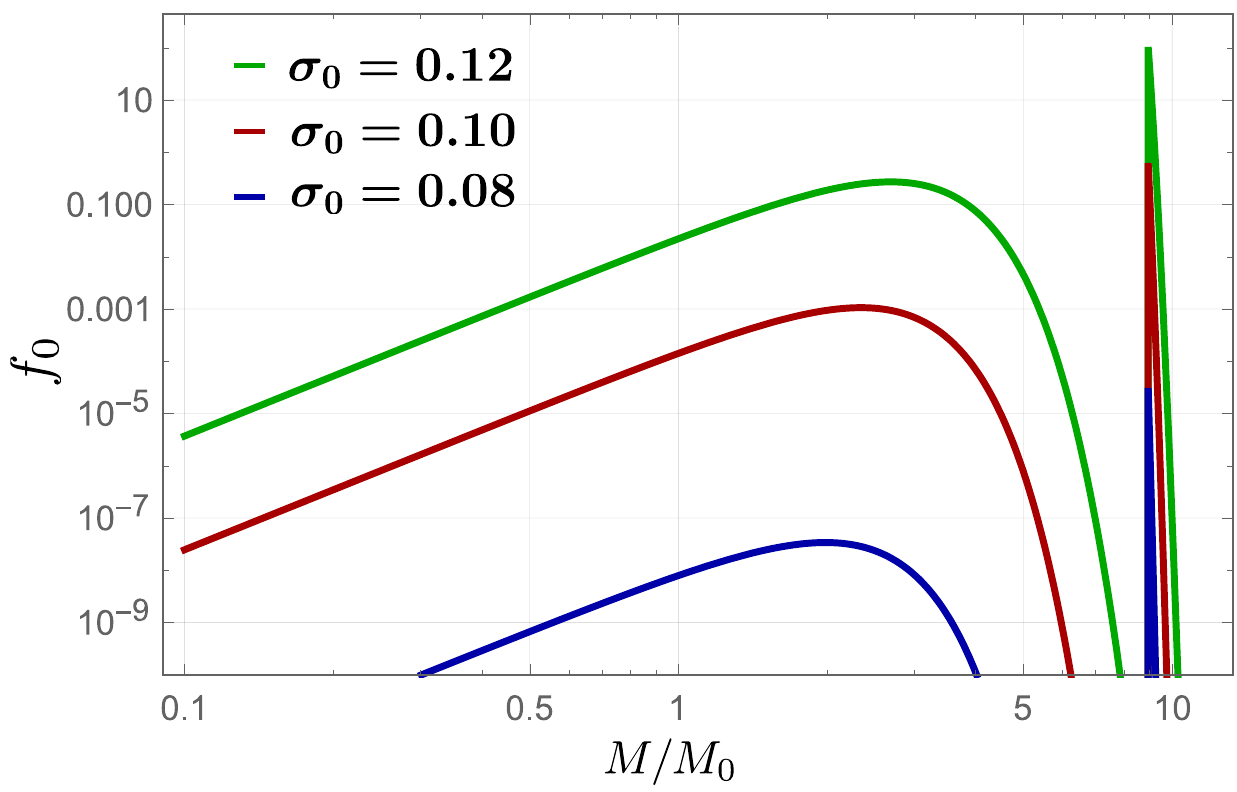}
\end{center}
\caption{The PBH fractions $f_0$ are plotted as functions of $M/M_0$ for $\sigma_0=0.08$, $0.10$ and $0.12$ with $k_0=10^5k_{\rm eq}$. 
The right spiky plots show the PBH fractions without the critical behavior, and the left broader plots show those with the critical behavior. 
}
\label{fig:f0}
\end{figure}

\begin{figure}[htbp]
\begin{center}
\includegraphics[width=12cm]{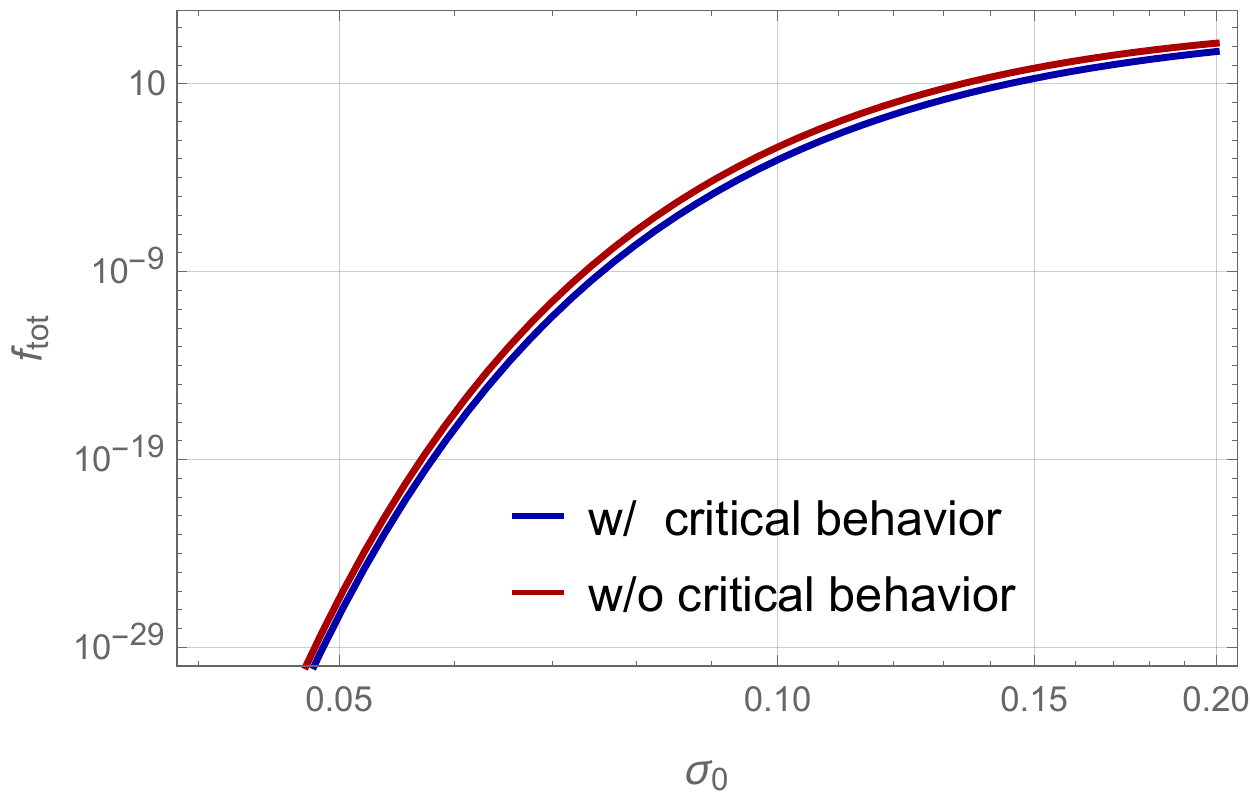}
\end{center}
\caption{The total fraction $f_{\rm tot}$ as a function of $\sigma_0$. }
\label{fig:ftot}
\end{figure}

For the number density, we reconsider the expression \eqref{eq:nkbullet-t}. 
In the limit $\gamma_3\rightarrow 0$, the probability distribution $P_1(\nu,\xi_1;\gamma_3)$ can be written as 
\begin{equation}
  \lim_{\gamma_3\rightarrow1}P_1(\nu,\xi_1;\gamma_3)=\lim_{\gamma_3\rightarrow1}\frac{2}{2\pi}\frac{1}{\sqrt{1-\gamma_3}}\left[-\frac{1}{2}\left(\nu^2+\frac{(\xi_1-\gamma_3\nu)^2}{1-\gamma_3^2}\right)\right]=\delta(\xi_1-\nu)\frac{1}{\sqrt{2\pi}}\exp\left[-\frac{1}{2}\nu^2\right]. 
\end{equation}
Substituting $\nu=\mu_2/\sigma_2$ and $\xi_1=\mu_2 k_\bullet^2/\sigma_4$ and using $\sigma_n=\sigma_0k_0$,  
we obtain 
\begin{equation}
  P_1\left(\frac{\tilde \mu_2}{\sigma_0},\frac{\tilde \mu_2 k_\bullet^2}{\sigma_0 k_0^2};1\right)=\frac{\sigma_0 k_0}{2\tilde \mu_2}\delta(k_\bullet -k_0)\frac{1}{\sqrt{2\pi}}\exp\left[-\frac{1}{2}\frac{\tilde \mu_2^2}{\sigma_0^2}\right]. 
\end{equation}
Integrating the number density for $k_\bullet$, we obtain 
\begin{equation}
n_{\rm pk}^{(\tilde \mu_2)}\dd \tilde \mu_2:=\int\dd k_\bullet  n_{\rm pk}^{(k_\bullet)}(\tilde \mu_2,k_\bullet) \dd \tilde \mu_2=
3^{-3/2}(2\pi)^{-2}\frac{1}{\sigma_0}k_0^3f\left(\frac{\tilde\mu_2}{\sigma_0}\right)\exp\left(-\frac{\tilde \mu_2^2}{2\sigma_0^2}\right). 
\end{equation}
The variable $\tilde \mu_2$ can be converted into $M$ through Eq.~\eqref{eq:Mmu}. 
Then we obtain 
\begin{eqnarray}
  n_{\rm BH}(M)d\ln M&=&n^{(\tilde \mu_2)}_{\rm pk} \left|\frac{\dd\ln M}{\dd\tilde \mu_2}\right|^{-1}\Theta(M-M_{\rm min}) \dd\ln M
  \cr
  &=&3^{-3/2}(2\pi)^{-2}\frac{1}{\sigma_0}k_0^3f\left(\frac{\tilde \mu_2}{\sigma_0}\right)\exp\left(-\frac{\tilde \mu_2^2}{2\sigma_0^2}\right)\left|\frac{\dd\ln M}{\dd\tilde \mu_2}\right|^{-1}\Theta(M-M_{\rm min}) \dd \ln M,~~ 
\end{eqnarray}
where $\tilde \mu_2$ is regarded as a function of $M$ through \eqref{eq:Mmu}, and 
\begin{equation}
  \left|\frac{\dd\ln M}{\dd\tilde \mu_2}\right|=\left|\frac{\gamma}{\tilde \mu_2-\tilde \mu_{\rm 2th}}-2g_{\rm m}\right|\simeq
  \left|\frac{\gamma}{\tilde \mu_2-0.615}-0.282\right|. 
\end{equation}
The PBH fraction $f_0$ is given by 
\begin{equation}
  f_0\simeq\frac{1}{3^{5/2}\pi}\frac{1}{\sigma_0}\frac{M}{M_{\rm eq}}\frac{k_0^3}{k_{\rm eq}^3} f\left(\frac{\tilde \mu_2}{\sigma_0}\right)
  \exp\left(-\frac{\tilde \mu_2^2}{\sigma_0^2}\right)\left|\frac{\gamma}{\tilde \mu_2-0.615}-0.282\right|^{-1}
  \Theta\left(M-M_{\rm min}\right). 
\end{equation}
Since small values of the PBH mass can be realized for $\tilde \mu_2 \sim \tilde \mu_{\rm 2th}$, 
the behavior of $f_0$ in the small PBH mass limit is given by $f_0\propto M^{1+\gamma^{-1}}$. 
We show the PBH fraction $f_0(M)$ for $k_0/k_{\rm eq}=10^{5}$ 
as a function of $M/M_0$ in Fig.~\ref{fig:f0} with $M_0=M_{\rm eq}k_{\rm eq}^2/k_0^2$. 
The total fraction $f_{\rm tot}=\int f_0 \dd \ln M$ is shown as functions of $\sigma_0$ for the cases with and without critical behavior 
in Fig.~\ref{fig:ftot}

\section{Summary and further related topics}

In this review, we have run through the basic picture of PBH formation, cosmological long-wavelength solutions, 
PBH formation criterion and estimation of PBH abundance in peak theory. 
In this last section, let us summarize and make notes on each topic. 

First, in Sec.~\ref{sec:three_zone}, we have considered the three-zone model. 
It should be noted that the three-zone model cannot be an exact solution for the Einstein field equations with non-vanishing pressure of the fluid, 
and just gives us intuition about the PBH formation process. 
Nevertheless, it is interesting that the three-zone model provides 
us a quantitative threshold value that is roughly equal to the lower bound of the threshold value of the density perturbation~\cite{Musco:2018rwt}. 
Similar ideas to the three-zone model would be very useful as a foothold when we extend our focus to PBH formations in other settings.

For more accurate analyses, the cosmological long-wavelength solutions are needed to be considered. 
In Sec.~\ref{sec:cosmo_long_sol}, we just quoted the equations and solutions from Refs.~\cite{Shibata:1999zs,Harada:2015yda}. 
Readers can refer to Refs.~\cite{Shibata:1999zs,Harada:2015yda} for more general and concrete derivations and discussions. 
We note that in Sec.~\ref{sec:cosmo_long_sol}, we have not assumed spherical symmetry. 
In this sense, the formulation given in Sec.~\ref{sec:cosmo_long_sol} is more general than the gradient expansion 
in the Misner-Sharp formulation (e.g., Ref.~\cite{Polnarev:2006aa}) which is widely used to construct the initial data for numerical simulations. 
When one considers other matter components than the perfect fluid with the linear equation of state $p=w\rho$, 
the cosmological long-wavelength solution should be reconsidered. For example, in Ref.~\cite{Yoo:2021fxs}, 
cosmological long-wavelength solutions for massless scalar isocurvature modes have been investigated. 

In Sec.~\ref{sec:PBH_formation_cri}, assuming spherical symmetry, we have considered the criterion of PBH formation. 
Although we need numerical simulations for an accurate criterion, 
some of the approximate analytic criteria are still practically useful. 
As a useful and relatively simple criterion, we have introduced the criterion based on the compaction function. 
The definition of the compaction function is carefully introduced in Sec.~\ref{sec:PBH_formation_cri} 
because we recently found a misunderstanding that has not been realized for a long time~\cite{Harada:2023ffo}. 
More concrete discussions will be reported elsewhere~\cite{Harada:2023ffo}. 
Since the compaction function is closely related to the conserved mass in a spherically symmetric spacetime, 
we have provided a short review of the conserved mass and horizons in spherically symmetric spacetimes in Appendix~\ref{sec:horizon_mass}. 

One of the goals of the theoretical PBH study is the estimation of PBH abundance. 
Since PBHs would be associated with rarely high peaks of a perturbation variable, 
the peak statistics are relevant to be used. 
We have attached a short minimal review of the peak theory in Appendix~\ref{sec:number_density} and \ref{sec:typical_profile}. 
We have explained the abundance estimation following Refs.~\cite{Yoo:2018kvb,Yoo:2020dkz} mainly focusing on the case of the monochromatic curvature power spectrum. 
However, different procedures based on peak theory have been also proposed~\cite{Germani:2018jgr,Germani:2019zez,Musco:2020jjb,Suyama:2019npc}.  
Here we note that, while the variation of the peak curvature scale is treated as an independent statistical variable in Refs.~\cite{Yoo:2018kvb,Germani:2019zez,Yoo:2020dkz}, the typical profile for the mean peak curvature is only considered in Refs.~\cite{Germani:2018jgr,Musco:2020jjb}. 
We may expect that those different procedures predict similar PBH mass spectra for a narrow curvature power spectrum. 
However, the predictions may be different from each other for a broad curvature power spectrum. 
This issue is still open and a direct comparison between those different procedures is needed to make a deeper understanding and a consensus. 

There are still many interesting open issues we could not mention here such as effects of non-Gaussianity\cite{Bullock:1996at,Ivanov:1997ia,Yokoyama:1998xd,Hidalgo:2007vk,Byrnes:2012yx,Bugaev:2013vba,Young:2015cyn,Nakama:2016gzw,Ando:2017veq,Franciolini:2018vbk,Cai:2018dig,Ando:2017veq,Atal:2018neu,Passaglia:2018ixg,Yoo:2019pma,Taoso:2021uvl,Atal:2019cdz,Atal:2019erb,Kitajima:2021fpq,Biagetti:2021eep,Escriva:2022pnz} PBH in matter dominated era\cite{Khlopov:1980mg,Polnarev:1985btg,Harada:2016mhb,Harada:2017fjm,Kokubu:2018fxy}, non-spherical PBH formation~\cite{Yoo:2020lmg}, distribution of PBH spin\cite{Chiba:2017rvs,DeLuca:2020bjf,Harada:2017fjm,Mirbabayi:2019uph,He:2019cdb,Flores:2021tmc,Koga:2022bij}, PBH formation in the QCD phase transition epoch~\cite{Jedamzik:1996mr,Schmid:1998mx,Widerin:1998my,Boeckel:2010bey,Sobrinho:2016fay,Byrnes:2018clq,Carr:2019kxo,Carr:2019hud,Clesse:2020ghq,Gao:2021nwz,Escriva:2022pnz,Franciolini:2022tfm,Juan:2022mir} and 
PBH from iso-curvature perturbations\cite{Passaglia:2021jla,Yoo:2021fxs}, and so on. 
PBH formation in modified gravity theories would require more careful treatments in all aspects. 
We hope this review will be helpful for researchers to study those issues in the future. 

\section*{Acknowledgment}
This review is based on the lecture given at Asia-Pacific School and Workshop on Gravitation and Cosmology 2022. 
This work was supported in part by JSPS KAKENHI Grant Numbers JP19H01895~, JP20H05850~ and JP20H05853.

\appendix
\section{Horizon and mass}
\label{sec:horizon_mass}

Throughout this section, we assume spherical symmetry for simplicity 
and review the references \cite{Misner:1964je,Kodama:1979vn,Hayward:1993wb,Hayward:1994bu}. 
In the case of the standard PBH formation scenario in RD, 
this assumption can be justified as will be discussed in Appendix~\ref{sec:number_density}.

First, let us write the line element of a spherically symmetric spacetime as follows:
\begin{equation}
  \dd s^2=-\ee^{2\phi(t,r)}\dd t^2+\ee^\lambda(t,r)\dd r^2+R^2(t,r)\dd \Omega^2. 
  \label{eq:diagmet}
\end{equation}
Introducing the null coordinates satisfying 
\begin{equation}
  \sqrt{2}\ee^{-f(t,r)}\dd \xi_\pm=\ee^\phi \dd t\pm \ee^{\lambda /2}\dd r, 
\end{equation}
we can rewrite the line element as 
\begin{equation}
  \dd s^2=  -2\ee^{-f(t,r)}\dd \xi_+ \dd \xi_-+R^2(t,r)\dd\Omega^2. 
\end{equation}

In this general spherically symmetric spacetime, 
existence of causal horizons, such as cosmological and black hole horizons, 
can be characterized by marginally trapped surfaces. 
Defining the null expansion $\theta_\pm$ as 
\begin{equation}
  \theta_{\pm}:=\frac{2}{R}\frac{\del R}{\del \xi_\pm}, 
\end{equation}
we can define a marginally trapped surface as a surface satisfying 
\begin{equation}
  \theta_+\theta_-=0. 
\end{equation}
For instance, assuming $\del_t$ and $\del_r$ as future-directed and outward vector fields, respectively, 
we find that $\theta_+=0$ for the black hole horizon in the Schwarzschild black hole and 
$\theta_-=0$ for the cosmological horizon in the spatially flat FLRW universe models. 

The definition of the Misner-Sharp mass is originally given by \cite{Misner:1964je}
\begin{equation}
  M(t,r)=\frac{1}{2}R\left[1+\ee^{-2\phi}(\del_t R)^2-\ee^{-\lambda}(\del_r R)^2\right]. 
  \label{eq:MSmass}
\end{equation}
As can be easily checked, we can also write the Misner-Sharp mass in the following different forms:
\begin{eqnarray}
  M(t,r)&=&\frac{1}{2}R\left(1-g^{\mu\nu}\del_\mu R\del_\nu R\right)\\
  &=&\frac{1}{2}R+\frac{1}{4}\ee^f R^3 \theta_+\theta_-. 
\end{eqnarray}
From this expression, we find 
\begin{equation}
  R(t,r)=2M(t,r)~~~\text{if}~~~\theta_+\theta_-=0. 
\end{equation}
That is, half of the horizon radius is equal to the Misner-Sharp mass enclosed by the horizon in general spherically symmetric spacetimes. 

It would not be easy to understand the expression \eqref{eq:MSmass} describes the mass of the system. 
In order to see the relation between the stress-energy tensor and the Misner-Sharp mass, 
let us write down the Einstein equations in terms of the Misner-Sharp mass and the 
stress-energy tensor. 
We obtain two independent equations as follows:
\begin{eqnarray}
  \del_r M&=&4\pi R^2\left(T^t_{~t}\del_rR-T^t_{~r}\del_tR\right), 
  \label{eq:delrM}\\
  \del_t M&=&4\pi R^2\ee^{-\lambda}\left(T^r_{~t}\del_rR-T^r_{~r}\del_tR\right). 
\end{eqnarray}
For instance, in the case of a perfect fluid, 
we can take the comoving coordinate for $t$ and $r$, that is, $T^t_{~r}=0$. 
Then, from Eq.~\eqref{eq:delrM}, we obtain 
\begin{equation}
  M=4\pi \int T^t_{~t}R^2 \dd R. 
  \label{eq:MSint}
\end{equation}
This expression would be much more familiar and understandable than Eq.~\eqref{eq:MSmass}. 
Nevertheless, the expression \eqref{eq:MSint} contains the integral and looks non-local quantity
in contrast to the expression \eqref{eq:MSmass}. 
Eq. \eqref{eq:MSmass} is often more useful than the expression \eqref{eq:MSint} because it is composed of only locally defined geometrical quantities 
and independent of the matter components. 

Another way to define the quasi-local mass in spherical symmetry associated with a conserved current has been proposed by Kodama~\cite{Kodama:1979vn}. 
The mass defined by Kodama is equivalent to the Misner-Sharp mass as will be shown soon. 
Let us define the following vector, often called Kodama vector, as follows:
\begin{equation}
  K^A:=-\epsilon^{AB}\del_BR, 
\end{equation}
where $A$ and $B$ run over $t$ and $r$, and $\epsilon^{AB}$ is the 2-dimensional totally anti-symmetric tensor given by 
\begin{equation}
  \epsilon_{AB}=\ee^{\phi+\lambda/2}\left(
  \begin{array}{cc}
    0&1\\-1&0
  \end{array}
  \right),~~
  \epsilon^{AB}=\ee^{-\phi-\lambda/2}\left(
    \begin{array}{cc}
      0&-1\\1&0
    \end{array}
    \right). 
\end{equation}
The sign on the right-hand side is a convention that makes the Kodama vector future-directed for $\del_r R>0$. 
This vector in the 2-dimensional spacetime spanned by $t$ and $r$ can be trivially 
promoted to the vector field in the 4-dimensional spacetime as 
\begin{equation}
  K^\mu=K^t\left(\frac{\del}{\del t}\right)^\mu+K^r\left(\frac{\del}{\del r}\right)^\mu. 
\end{equation}
Then the Kodama current $S^\mu$ is defined by 
\begin{equation}
  S^\mu:=-T^\mu_{~\nu}K^\nu, 
  \label{eq:Kcur}
\end{equation}
where we have fixed the sign on the right-hand side such that $S^t>0$ for $K^t>0$.

Let us check that the Kodama current is conserved. 
By using the Einstein equations, we can rewrite the Kodama current \eqref{eq:Kcur} as follows:
\begin{eqnarray}
  S^A=-T^A_{~B}K^B&=&\ee^{-\phi-\lambda/2}\frac{1}{4\pi R^2}\left[\del_r M(\del_r)^A-\del_t M(\del_r)^A\right]\cr
  &=&-\frac{1}{4\pi R^2}\epsilon^{AB}\del_B M. 
\end{eqnarray}
Then we obtain 
\begin{equation}
  \nabla_\mu S^\mu=\frac{1}{R^2\ee^{\phi+\lambda/2}}\del_A(R^2\ee^{\phi+\lambda/2}S^A)=-\frac{1}{4\pi R^2}\frac{\epsilon^{AB}}{\ee^{\phi+\lambda/2}}\del_A\del_BM=0, 
\end{equation}
where we have used $\del_\mu (\epsilon^{AB}/\ee^{\phi+\lambda/2})=0$. 
Therefore we can define the following conserved charge (Kodama mass):
\begin{eqnarray}
  M_{\rm K}&:=&-\int S^\mu n_\mu \dd \Sigma
  \label{eq:Kodama_mass}\\
  &=&\frac{1}{4\pi}\int \frac{1}{R^2}\epsilon^{AB}\del_B M n_A \dd \Sigma=\int \del_r M \dd r =M, 
\end{eqnarray}
where $\dd\Sigma$ is the volume element of the $t=\text{const.}$ hypersurface whose unit normal form is given by $n_\mu=-\ee^\phi(\dd t)_\mu$. 
This equation explicitly shows the equivalence between the Kodama mass and the Misner-Sharp mass. 

\section{Peak number density in peak theory}
\label{sec:number_density}

In this and the next section, we briefly review the necessary part of the peak theory~\cite{1986ApJ...304...15B} for the calculation of PBH abundance.  
Let us consider the Gaussian variable with the power spectrum $\mathcal P(k)$ defined by 
\begin{equation}
  \langle\tilde \zeta^*(\bm k)\tilde \zeta(\bm k')\rangle=\frac{2\pi^2}{k^3}\mathcal P(k)(2\pi)^3\delta(\bm k-\bm k')
\end{equation}
with $\tilde \zeta(\bm k)=\int \dd^3 x \zeta(\bm x)\ee^{\ii \bm k \bm x}$. 
The gradient moments $\sigma_n$ are given by 
\begin{equation}
  \sigma_n^2=\int \dd \ln k k^{2n}\mathcal P(k). 
\end{equation}
The probability distribution of linear combinations of $\zeta(\bm x)$ denoted by $V_I~(I=1,2,\cdots,n)$ is given by 
the multivariate Gaussian distribution:
\begin{equation}
  P(V_I)\dd^n V=(2\pi)^{-n/2}\left|\det \mathcal M\right|^{-1/2}\exp\left[-\frac{1}{2}V_I (\mathcal M^{-1})^{IJ} V_J \right] \dd^n V, 
\end{equation}
where 
\begin{equation}
  \mathcal M_{IJ}=\int \frac{\dd^3 k}{(2\pi)^3}  \frac{\dd^3 k'}{(2\pi)^3} \langle\tilde V^*_I(\bm k)\tilde V_J(\bm k')\rangle.  
\end{equation}
For example, let us consider the probability distribution of $-\zeta$ and $\triangle \zeta$ at a spatial point, 
where $\triangle$ is the flat Laplacian. 
The correlation matrix can be calculated as 
\begin{equation}
  \mathcal M_{IJ}=
  \left(
    \begin{array}{cc}
      \sigma_0^2&\sigma_1^2\\
      \sigma_1^2&\sigma_2^2
    \end{array}
  \right). 
\end{equation}
Then the probability distribution of $-\zeta$ and $\triangle \zeta$ is given by 
\begin{eqnarray}
  &&P(-\zeta,\triangle \zeta)\dd(-\zeta)\dd(\triangle \zeta)=\frac{1}{2\pi\sqrt{\sigma_0^2\sigma_2^2-\sigma_1^4}}\cr
  &&\hspace{1cm}\times\exp\left[-\frac{1}{2(\sigma_0^2\sigma_2^2-\sigma_1^4)}\left\{\sigma_0^2(-\zeta)^2+\sigma_2^2(\triangle\zeta)^2-2\sigma_1^2(-\zeta)(\triangle \zeta)\right\}\right]\dd(-\zeta)\dd(\triangle \zeta). 
\end{eqnarray}

Similarly to the example of $-\zeta$ and $\triangle \zeta$, let us consider 
the probability distribution function of the coefficients in Taylor expansion of $\zeta$ up through the second order: 
\begin{equation}
  \zeta=\zeta_0+\sum_i\zeta^i_1 x_i+\sum_{i,j}\frac{1}{2}\zeta_2^{ij}x_ix_j+\mathcal O(x^3). 
\end{equation}
There are 10 independent variables $\zeta_0$, $\zeta_1^i~(i=1,2,3)$ and $\zeta_2^{ij}~(i,j=1,2,3)$. 
Non-zero correlations between two of these coefficients are summarized as follows:
\begin{eqnarray}
  \sigma_0^2&=&\int \dd \ln k\mathcal P(k)=\langle\zeta_0\zeta_0\rangle,\\
  \sigma_1^2&=&\int \dd \ln k k^2\mathcal P(k)=-3\langle\zeta_0\zeta_2^{ii}\rangle=3\langle\zeta_1^i\zeta_1^i\rangle,\\
  \sigma_2^2&=&\int \dd \ln k k^4\mathcal P(k)=-5\langle\zeta_2^{ii}\zeta_2^{ii}\rangle=15\langle\zeta_2^{ii}\zeta_2^{jj}\rangle=15\langle\zeta_2^{ij}\zeta_2^{ij}\rangle ~{\rm with}~i\neq j. 
\end{eqnarray}
The 2nd order variables can be transformed into the 
six independent variables $\lambda_i$ and $\theta_i$ with $i=1,2,3$, where 
$\lambda_1$, $\lambda_2$ and $\lambda_3$ are the three eigenvalues 
of the matrix $\zeta_2^{ij}$ with $\lambda_1\geq\lambda_2\geq\lambda_3$ 
and $\theta_1$, $\theta_2$ and $\theta_3$ are the Euler angles to take the principal direction. 
Then it can be shown that the volume element $\dd^6\zeta_2$ can be given by 
\begin{equation}
  \dd^6\zeta_2=(\lambda_1-\lambda_2)(\lambda_2-\lambda_3)(\lambda_1-\lambda_3)\dd^3\lambda \sin\theta_1\dd^3\theta. 
\end{equation}
The integration with respect to the Euler angles gives the factor $2\pi$. 

We introduce the following 7 independent variables by using the remaining 7 variables:
\begin{eqnarray}
  \nu&=&-\frac{\zeta_0}{\sigma_0},\\
  \eta_i&=&\frac{\zeta^i_1}{\sigma_1}, \\
  \xi_1&=&\frac{\lambda_1+\lambda_2+\lambda_3}{\sigma_2},\\
  \xi_2&=&\frac{1}{2}\frac{\lambda_1-\lambda_3}{\sigma_2},\\
  \xi_3&=&\frac{1}{2}\frac{\lambda_1-2\lambda_2+\lambda_2}{\sigma_2}. 
\end{eqnarray}
Then the probability distribution is transformed into the following form:
\begin{equation}
  P(\nu,\bm \xi,\bm \eta)\dd \nu\dd^3\xi\dd^3\eta=P_1(\nu,\xi_1;\gamma_1)P_2(\xi_2,\xi_3)P_3(\bm \eta)\dd \nu\dd^3\xi \dd^3\eta, 
\end{equation}
where 
\begin{eqnarray}
  P_1(\nu,\xi_1;\gamma_1)\dd \nu\dd\xi_1&=&\frac{1}{2\pi}\frac{1}{\sqrt{1-\gamma_1^2}}\exp\left[-\frac{1}{2}\left(\nu^2+\frac{(\xi_1-\gamma_1\nu)^2}{1-\gamma_1^2}\right)\right]\dd\nu\dd\xi_1,
  \label{eq:p1}
  \\
P_2(\xi_2, \xi_3)\dd\xi_2\dd\xi_3&=&\frac{5^{5/2}3^2}{(2\pi)^{1/2}}\xi_2(\xi_2^2-\xi_3^2)\exp\left[-\frac{5}{2}(3\xi_2^2+\xi_3^2)\right]\dd \xi_2\dd\xi_3, \\
P_3(\bm \eta)\dd^3\eta&=&\frac{3^{3/2}}{(2\pi)^{3/2}}\exp\left[-\frac{3}{2}\left(\eta_1^2+\eta_2^2+\eta_3^2\right)\right]\dd^3 \eta
\end{eqnarray}
with 
\begin{equation}
  \gamma_n=\frac{\sigma_n^2}{\sigma_{n-1}\sigma_{n+1}}. 
\end{equation}
It would be worthy to note that there are no correlations between $(\nu,\xi_1)$ and $(\xi_2,\xi_3)$. 
Therefore even if we consider a rarely high peak with $\nu\sim\xi_1\gg 1$, the values of $\xi_2$ and $\xi_3$ are 
expected to be $\sim \mathcal O(1)\ll\xi_1$. 
Since $\xi_2$ and $\xi_3$ are responsible for the asphericity of the peak, 
we can conclude that rarely high peaks tend to be more spherically symmetric.


Let us remember that what we are seeking is the mean number density of extrema characterized by the amplitude and curvature of the peak. 
The corresponding variables to the amplitude and the peak curvature would be $\nu$ and $\xi_1$. 
Therefore our goal in this section is to obtain the mean number density characterized by the parameters $\nu$ and $\xi_1$: $\bar n_{\rm peak}(\nu,\xi_1)$. 
As a first step, let us consider the number density distribution of extrema $n_{\rm ext}(\bm x,\nu,\xi_1)$ in the space spanned by $(\bm x,\nu,\xi_1)$. 
By definition, we have 
\begin{eqnarray}
  n_{\rm ext}(\bm x,\nu,\xi_1)\Delta \bm x \Delta \nu \Delta \xi_1&=&\text{number of extrema in }\Delta \bm x \Delta \nu \Delta \xi_1\cr
  &=&\sum_p\delta(\bm x-\bm x_p)\delta(\nu-\nu_p)\delta(\xi_1-\xi_{1p})\Delta \bm x\Delta \nu \Delta \xi_1, 
\end{eqnarray}
where $p$ labels the extrema so that the $p$-th extremum will be characterized by $(\bm x_p, \nu_p, \xi_{1p})$. 
The integration over the space spanned by $(\bm x,\nu,\xi_1)$ trivially gives the total number of extrema. 
This expression of the number density distribution is the specific distribution for a given set of the parameters $(\bm x_p, \nu_p, \xi_{1p})$. 
Since we are interested in the average distribution without the explicit dependence on $(\bm x_p, \nu_p, \xi_{1p})$, 
let us take the volume average and the ensemble average for $\nu_p$ and $\xi_{1p}$ of the distribution as follows:
\begin{eqnarray}
  \bar n_{\rm ext}(\nu,\xi_1)\Delta \nu \Delta \xi_1&:=&\frac{1}{V}\int \dd^3 x \prod_p \int \dd\nu_p \dd \xi_{1p} P_1(\nu_p,\xi_{1p};\gamma_1)
  n_{\rm ext}(\bm x,\nu,\xi_1)\Delta\nu \Delta \xi_1 \cr
  &=&P_1(\nu,\xi_1;\gamma_1)\Delta \nu \Delta \xi_1 \frac{1}{V}\int \dd^3 x \sum_p\delta(\bm x-\bm x_p). 
\end{eqnarray}
To dispose the $\delta(\bm x -\bm x_p)$ and the volume integral, let us consider rewriting the delta function. 
Since we are focusing on extrema, at $\bm x=\bm x_p$, $\bm \eta$ vanishes. 
Therefore we find 
\begin{eqnarray}
  &&\delta(\bm \eta (\bm x))=\sum_p \left|\det\left(\frac{\del \bm\eta}{\del \bm x}\right)\right|_p^{-1}\delta(\bm x-\bm x_p)=\left|\det\left(\frac{\del \bm\eta}{\del \bm x}\right)\right|^{-1}\sum_p \delta(\bm x-\bm x_p)\\
&\Leftrightarrow&\sum_p \delta(\bm x-\bm x_p)=\sigma_1^{-3}\left|\det \zeta_2^{ij}(\bm x)\right|\delta(\bm \eta)=\sigma_1^{-3}\left|\lambda_1\lambda_2\lambda_3\right|\delta(\bm\eta), 
\end{eqnarray}
where $\left|\lambda_1\lambda_2\lambda_3\right|$ can be written as 
\begin{equation}
  \left|\lambda_1\lambda_2\lambda_3\right|=\frac{\sigma_2^3}{27}\left|\left((\xi_1+\xi_2)^2-9\xi_2^2\right)\left(\xi_1-2\xi_3\right)\right|
\end{equation}
in terms of $\bm \xi$. 
Then we obtain
\begin{equation}
  \bar n_{\rm ext}(\nu,\xi_1)=P_1(\nu,\xi_1;\gamma_1)\frac{\sigma_2^3}{3^3\sigma_1^3}\frac{1}{V}\int \dd \bm x \left|\left((\xi_1+\xi_2(\bm x))^2-9\xi_2^2(\bm x)\right)\left(\xi_1-2\xi_3(\bm x)\right)\right| \delta(\bm \eta(\bm x)),  
\end{equation}
where we have fixed the value of $\xi_1$ and take $\xi_1$ as an independent variable of $\bm x$. 
Here we replace the volume average with the ensemble average for $\xi_2$, $\xi_3$ and $\bm \eta$ as follows:
\begin{eqnarray}
  &&\bar n_{\rm ext}(\nu,\xi_1)=P_1(\nu,\xi_1;\gamma_1)\frac{\sigma_2^3}{3^3\sigma_1^3}\int\dd \xi_2\dd\xi_3\dd^3 \eta P_2(\xi_2,\xi_3)P_3(\bm \eta) \left|\left((\xi_1+\xi_2)^2-9\xi_2^2\right)\left(\xi_1-2\xi_3\right)\right| \delta(\bm \eta)\cr
  &&~~=    P_1(\nu,\xi_1;\gamma_1)\frac{\sigma_2^3}{3^3\sigma_1^3}P_3(\bm \eta=0)\int\dd \xi_2\dd\xi_3 P_2(\xi_2,\xi_3) \left|\left((\xi_1+\xi_2)^2-9\xi_2^2\right)\left(\xi_1-2\xi_3\right)\right|. 
\end{eqnarray}
The peak number density can be obtained by multiplying the step function $\Theta(\lambda_3)$. 
Then the integration for $\xi_2$ and $\xi_3$ can be explicitly done~\cite{1986ApJ...304...15B} and we obtain
\begin{equation}
  n_{\rm peak}(\nu,\xi_1)=3^{-3/2}(2\pi)^{-3/2}\frac{\sigma_2^3}{\sigma_1^3}f(\xi_1)P_1(\nu,\xi_1;\gamma_1), 
\end{equation}
where 
\begin{eqnarray}
  f(\xi_1)&:=&\frac{1}{2}\xi_1(\xi_1^2-3)
\left({\rm erf}\left[\frac{1}{2}\sqrt{\frac{5}{2}}\xi_1\right]
+{\rm erf}\left[\sqrt{\frac{5}{2}}\xi_1\right]\right)
\cr
&&+\sqrt{\frac{2}{5\pi}}\left\{
\left(\frac{8}{5}+\frac{31}{4}\xi_1^2\right)
\exp\left[-\frac{5}{8}\xi_1^2\right]
+\left(-\frac{8}{5}+\frac{1}{2}\xi_1^2\right)\exp\left[-\frac{5}{2}\xi_1^2\right]
\right\}. 
\label{eq:funcf}
\end{eqnarray}
For convenience in the calculation of PBH abundance, let us change the variables from $(\xi_1,\nu)$ 
to $(\mu_0,k_*)$ defined by 
\begin{eqnarray}
  \mu_0&=&-\zeta_0=\sigma_0\nu,\\
  \mu_0 k_*^2&=&\triangle\zeta|_{\rm peak}=\sigma_2\xi_1. 
\end{eqnarray}
Then the peak number density can be transformed into the following form:
\begin{equation}
  n^{(k_*)}_{\rm pk}(\mu_0, k_*) d \mu_0 d k_*
:=n_{\rm peak }(\nu,\xi_1)d \nu d\xi_1
=\frac{2 \mu_0 k_*}{3^{3/2}(2\pi)^{3/2}}\frac{\sigma_2^2}{\sigma_0\sigma_1^3}f\left(\frac{\mu_0 k_*^2}{\sigma_2}\right)
P_1\left(\frac{\mu_0}{\sigma_0},\frac{\mu_0 k_*^2}{\sigma_2};\gamma_1\right) 
d\mu_0 d k_*. 
\end{equation}

\section{Typical profile in peak theory}
\label{sec:typical_profile}

The peak theory also provides a typical profile characterized by the peak parameter $\nu$ and $\xi_1$~\cite{1986ApJ...304...15B}. 
To derive the expression of the typical profile, let us consider the conditional probability of 
taking the value $\zeta(r)$ at the radius $r$ from a peak. 
Before going into details, we define some notations about conditional probabilities. 
We represent the probability for the realization of the event $X$ as $P_X$. 
The conditional probability for the realization of $Y$ under the condition $X$ is denoted by $P_{Y|X}$ and given by 
\begin{equation}
  P_{Y|X}=\frac{P_{X\cap Y}}{P_X}, 
  \label{eq:condprob}
\end{equation} 
where $P_{X\cap Y}$ is the probability for the realization of $X$ and $Y$. 
Let us consider the multi-variate Gaussian probability for the probability variables given by 
\begin{equation}
  q_\mu=(q_1,q_2,\cdots, q_N)=(\alpha_1,\cdots, \alpha_n,\beta_1,\cdots, \beta_{N-n}), 
\end{equation}
where $\alpha_a~(a=1,\cdots, n)$ and $\beta_i~(i=1,\cdots, N-n)$ correspond to the variables for $X$ and $Y$, respectively. 
Letting $q^\mu$ denotes the components of the transposition of the row vector $q$ as $q^\mu=(q^{\rm t})^\mu$, 
the probability distribution of $X\cup Y$ can be written as 
\begin{equation}
  P_{X\cup Y}(q_\mu)=\frac{1}{(2\pi)^{N/2}(\det M)^{1/2}}\exp\left[-\frac{1}{2}\sum_{\mu,\nu} q_\mu \left(M^{-1}\right)^\mu_{~\nu}q^\nu\right], 
\end{equation}
where $M$ is the $N\times N$ correlation matrix. 
We decompose the correlation matrix in the following form:
\begin{equation}
  M=\left(\begin{array}{cc}
    A^a_{~b}&B^a_{~j}\\
    C^i_{~b}&D^i_{~j}
  \end{array}\right). 
\end{equation}
Then the inverse matrix of $M$ is given by 
\begin{equation}
  M^{-1}=
\left(
\begin{array}{cc}
A^{-1}+A^{-1}B(M/A)^{-1}CA^{-1}&-A^{-1}B(M/A)^{-1}\\
-(M/A)^{-1}CA^{-1}&(M/A)^{-1}
\end{array}
\right), 
\end{equation}
where $M/A$ is the Schure complement of the block $A$ given by 
\begin{equation}
  M/A=D-CA^{-1}B. 
\end{equation}
By using the expression of $M^{-1}$, we find 
\begin{equation}
  q_\mu\left(M^{-1}\right)^\mu_{~\nu}q^\nu
  =\alpha A^{-1}\alpha^{\rm t}+(\beta-\alpha A^{-1}B)(M/A)^{-1}(\beta-\alpha A^{-1}B)^{\rm t}, 
\end{equation}
where we have used the fact $B=C^{\rm t}$. 
Since the conditional probability is given by Eq.~\eqref{eq:condprob}, 
we find 
\begin{equation}
  P_{Y|X}=\frac{1}{(2\pi)^{(N-n)/2}(\det M/A)^{1/2}} \exp\left[-\frac{1}{2}(\beta-\alpha A^{-1}B)(M/A)^{-1}(\beta-\alpha A^{-1}B)^{\rm t}\right]. 
\end{equation}
Therefore the mean values and the correlation matrix for $\beta$ under the condition $X$ are given by $\alpha A^{-1}B$ and $M/A$. 

We are interested in the probability of $\beta=\nu(r)=\zeta(r)/\sigma_0$ under the condition $\alpha_a=(\nu(0),\xi_1(0),\bm \eta(0)=0)$. 
To simplify the expressions we write $\nu(0)$, $\xi_1(0)$ and  $\bm \eta(0)$ without the argument as $\nu$, $\xi_1$ and  $\bm \eta$, 
but the argument of $\nu(r)$ will be kept. 
Then the 4 blocks of the correlation matrix can be calculated as  
\begin{eqnarray}
  A^a_{~b}&=&<\alpha^a\alpha^b>=\left(
\begin{array}{ccc}
1&\gamma_1&0\\
\gamma_1&1&0\\
0&0&\frac{1}{3}I_{3\times 3}
\end{array}\right),\\
B^a_{~1}&=&C^1_{~a}=\langle \alpha^a\nu(r)\rangle,\\
D&=&\langle \nu(r)\nu(r)\rangle=1. 
\end{eqnarray}
Therefore we obtain
\begin{eqnarray}
  \left.\left(\alpha A^{-1}B\right)\right|_{\nu,\xi_1,\bm \eta=0}&=&\frac{1}{1-\gamma_1^2}(\nu-\xi_1\gamma_1)\langle \nu\nu(r)\rangle+\frac{1}{\gamma_1^2}(-\nu\gamma_1+\xi_1)\langle\xi_1\nu(r)\rangle,\\
M/A=D-CA^{-1}B&=&1-\frac{1}{1-\gamma_1^2}\langle\nu\nu(r)\rangle^2+\frac{2\gamma_1}{1-\gamma_1^2}\langle\xi_1\nu(r)\rangle\langle\nu\nu(r)\rangle
\cr&&-\frac{1}{1-\gamma_1^2}\langle\xi_1\nu(r)\rangle^2-3\sum_{i=1}^3\langle\eta_i\nu(r)\rangle^2. 
\end{eqnarray}
The correlations are calculated as 
\begin{eqnarray}
  \langle\nu\nu(r)\rangle&=&\int \frac{\dd^3 k}{(2\pi)^3}\frac{\dd^3 k'}{(2\pi)^3}\langle\tilde \nu^*(\bm k)e^{-i\bm k\cdot \bm r}~\tilde \nu(\bm k')\rangle
  =\sigma_0^{-2}\int \frac{\dd^3 k}{(2\pi)^3}\frac{\dd^3 k'}{(2\pi)^3} e^{-i\bm k\cdot \bm r}
  \langle\tilde \zeta(\bm k)\tilde \zeta(\bm k')\rangle\cr
  &=&\sigma_0^{-2}\int\frac{\dd^3 k}{4\pi}e^{-i\bm k\cdot \bm r}\frac{1}{k^3}\mathcal P(k)
  =\sigma_0^{-2}\int \dd\ln k\frac{\sin(kr)}{kr}\mathcal P(k)=:\psi_0(r),\\
\langle\xi_1\nu(r)\rangle&=&-\frac{\sigma_0}{\sigma_2}\triangle \psi_0(r),\\
\sum_{i=1}^3\langle\eta_i\nu(r)\rangle^2&=&\frac{1}{\sigma_1^2}\left(\del_r\psi_0(r)\right)^2. 
\end{eqnarray}
Then the mean profile is given by 
\begin{equation}
  \langle\nu(r)|\nu,\xi_1,\bm \eta=0\rangle=\frac{\nu}{1-\gamma_1^2}\left(\psi+\frac{1}{3}R_1^2\triangle\psi_0\right)
  -\frac{\xi_1}{\gamma_1(1-\gamma_1^2)}\left(\gamma_1^2\psi+\frac{1}{3}R_1^2\triangle\psi_0\right), 
\end{equation}
where 
\begin{equation}
  R_n^2=\frac{3\sigma_n^2}{\sigma_{n+1}^2}. 
  \label{eq:Rn}
\end{equation}
The variance is given by
\begin{eqnarray}
  \langle(\Delta\nu(r))^2|\nu,\xi_1,\bm \eta=0\rangle&=&1-\frac{1}{1-\gamma_1^2}\psi_0^2-\frac{1}{\gamma_1^2(1-\gamma_1^2)}\left(2\gamma_1^2\psi_0+\frac{R_1^2}{3}\triangle\psi_0\right)\frac{R_1^2}{3}\triangle\psi_0\cr
  &&-\frac{5}{\gamma_1^2}\left(\frac{\del_r\psi_0}{r}-\frac{\triangle\psi_0}{3}\right)^2R_1^4-\frac{R_1^2}{\gamma_1^2}(\del_r \psi_0)^2. 
\end{eqnarray}
The variance is roughly estimated as $\Delta \nu(r)\sim\Delta\zeta/\sigma_0=\mathcal O(1)$, that is, $\Delta\zeta\sim \sigma_0$. 
Therefore the typical profile is a good approximation as long as $\zeta(r)\gg\sigma_0$ for a rarely high peak. 
Changing the independent variables from $(\nu,\xi_1)$ to $(\mu_0,k_*)$, we obtain the following form of the typical profile of $\bar \zeta(r)$:  
\begin{equation}
  \frac{\langle\zeta(r)|\mu_0,k_*,\bm \eta=0\rangle}{\mu_0}=-\frac{1}{1-\gamma_1^2}\left(\psi_0+\frac{1}{3}R_1^2\triangle\psi_0\right)+k_*^2\frac{1}{\gamma_1(1-\gamma_1^2)}\frac{\sigma_0}{\sigma_2}\left(\gamma_1^2\psi_0+\frac{1}{3}R_1^2\triangle\psi_0\right). 
\end{equation}

\section{Equations for peaks of $\triangle \zeta$}
\label{sec:peaksoflapzeta}

In practice, peaks of $\triangle \zeta$ are more important for the calculation of PBH abundance. 
Thus we consider the parameters $\mu_2=\triangle \zeta(0)$ and $k_\bullet^2:=-\frac{\triangle\triangle\zeta(0)}{\mu_2}$ instead of $\mu_0$ and $k_*^2$. 
With this modification associated with $-\zeta\rightarrow \triangle \zeta$, to obtain the number density and typical profile of peaks of $\triangle \zeta$, 
we need the following replacements:
\begin{equation}
  \psi_0\rightarrow \psi_2,~\mu_0\rightarrow\mu_2,~k_*\rightarrow k_\bullet,~\sigma_n\rightarrow \sigma_{n+2}~(\gamma_1\rightarrow\gamma_3,~R_1\rightarrow R_3), 
\end{equation}
where \footnote{The definition of $\psi_2$ is different from that in Refs.~\cite{Yoo:2020dkz,Yoo:2018kvb}, but the same as that in Ref.~\cite{Kitajima:2021fpq}.}
\begin{equation}
  \psi_n:=\frac{1}{\sigma_n}\int\dd \ln k ~k^{2n}\frac{\sin(kr)}{kr}\mathcal P(k).
  \label{eq:psin}
\end{equation}

Then, introducing the dimensionless quantity $\tilde \mu_2=\mu_2\sigma_1^2/\sigma_2^2$, we obtain the number density and the typical profile as follows:
\begin{eqnarray}
  n^{(k_\bullet)}_{\rm pk}(\tilde \mu_2, k_\bullet) d \tilde \mu_2 d k_\bullet
=\frac{2 \tilde \mu_2 k_\bullet}{3^{3/2}(2\pi)^{3/2}}\frac{\sigma_2^3\sigma_4^2}{\sigma_1^4\sigma_3^3}f\left(\tilde \mu_2 k_\bullet^2\frac{\sigma_2^2}{\sigma_1^2\sigma_4}\right)
P_1\left(\tilde \mu_2\frac{\sigma_2}{\sigma_1^2},\tilde \mu_2 k_\bullet^2\frac{\sigma_2^2}{\sigma_1^2\sigma_4};\gamma_3\right) 
d\tilde \mu_2 d k_\bullet, &&
\label{eq:nkbullet}\\
\frac{\overline{\triangle \zeta}(r)}{\tilde \mu_2}=\frac{1}{1-\gamma_3^2}\left(\psi_2+\frac{1}{3}R_3^2\triangle\psi_2\right)-k_\bullet^2\frac{1}{\gamma_3(1-\gamma_3^2)}\frac{\sigma_2}{\sigma_4}\left(\gamma_3^2\psi_2+\frac{1}{3}R_3^2\triangle\psi_2\right). ~&&
\label{eq:bar_delta_zeta}
\end{eqnarray}
The profile of $\bar \zeta$ associated with $\overline{\triangle \zeta}$ can be obtained by integrating \eqref{eq:bar_delta_zeta} as follows: 
\begin{equation}
  \frac{\overline{\zeta}(r)}{\tilde \mu_2}=-\frac{1}{1-\gamma_3^2}\left(\psi_1+\frac{1}{3}R_3^2\triangle\psi_1\right)+k_\bullet^2\frac{1}{\gamma_3(1-\gamma_3^2)}\frac{\sigma_2}{\sigma_4}\left(\gamma_3^2\psi_1+\frac{1}{3}R_3^2\triangle\psi_1 \right)+\frac{\sigma_2}{\sigma_1^2}\zeta_\infty.
  \label{eq:zetabar} 
\end{equation}
It can be shown that $\zeta_\infty$ can be also regarded as a Gaussian probability variable (see Ref.~\cite{Yoo:2020dkz}).


\end{document}